\documentclass{amsart}
\usepackage[foot]{amsaddr}
\usepackage[margin=1in]{geometry}
\usepackage{amsmath, amsfonts, amssymb}
\usepackage{color}
\usepackage{algpseudocode}
\usepackage{algorithm}
\usepackage{siunitx}
\usepackage{graphicx}
\usepackage{subcaption}
\usepackage{physics}
\usepackage{bbm}
\usepackage{tikz}
\usepackage{mathtools}
\usepackage{mathrsfs}
\usepackage{hyperref}
\usepackage{cleveref}
\usepackage{mathdots}

\newcommand{\ii}{\mathrm{i}}
\newcommand{\id}{\mathrm{id}}
\newcommand{\sigmaz}{\hat{\sigma}_z}
\newcommand{\sigmax}{\hat{\sigma}_x}
\newcommand{\dt}{{\Delta t}}
\newcommand{\dk}{{\Delta k}}
\newcommand{\e}{\mathrm{e}}
\newcommand{\SPAN}{\mathrm{span}}

\newcommand{\sgn}{\mathrm{sgn}}

\newcommand{\U}[2]{U_{\sigma_{#1}^\pm,\sigma_{#2}^\pm}^{(#1,#2)}}
\newcommand{\G}[2]{G_{\sigma_{#1}^\pm,\sigma_{#2}^\pm}}
\newcommand{\F}[2]{F_{\sigma_{#1}^\pm,\sigma_{#2}^\pm}^{(#1,#2)}}
\newcommand{\M}[2]{M_{\sigma_{#1}^\pm,\sigma_{#2}^\pm}^{(#1,#2)}}
\newcommand{\A}[2]{A_{\sigma_{#1}^\pm,\sigma_{#2}^\pm}^{(#1,#2)}}
\newcommand{\Ucal}[2]{\mathcal{U}_{{\sigma}(#2:#1)}^{(#1,#2)}}
\newcommand{\Ucall}[2]{\mathcal{U}_{{\varsigma}(#2:#1)}^{(#1,#2)}}
\newcommand{\Mcal}[2]{\mathcal{M}_{{\sigma}(#2:#1)}^{(#1,#2)}}
\newcommand{\Mscr}[3]{\mathscr{M}_{{\sigma}(#2:#1)}^{(#1,#2),[#3]}}

\definecolor{blue_violet}{rgb}{0.54, 0.17, 0.89}
\newcommand\zc[1]{{\color{red} [ZC: #1]}}
\newcommand\ki[1]{{\color{blue_violet} [ki: #1]}}

\theoremstyle{remark}
\newtheorem*{remark}{Remark}

\title{Tree-Based Implementation of the Small Matrix Path Integral for System-Bath Dynamics}

\author{Geshuo Wang}
\address[Geshuo Wang]{Department of Mathematics, National University of Singapore,
  Level 4, Block S17, 10 Lower Kent Ridge Road, Singapore 119076}
\email{geshuowang@u.nus.edu}

\author{Zhenning Cai}
\address[Zhenning Cai]{Department of Mathematics, National University of Singapore,
  Level 4, Block S17, 10 Lower Kent Ridge Road, Singapore 119076}
\email{matcz@nus.edu.sg}

\thanks{Zhenning Cai's work was supported by the Academic Research Fund of the Ministry of Education of Singapore under grant A-0004592-00-00.}


\begin{document}

\maketitle

\begin{abstract}
The small matrix path integral (SMatPI) method is an efficient numerical approach to simulate the evolution of a quantum system coupled to a harmonic bath. 
The method relies on a sequence of kernel matrices that defines the non-Markovian dynamics of the quantum system.
In the original SMatPI method, these kernels are computed indirectly through the QuAPI method.
Instead, we focus on the definition of the kernel matrices and reveal a recurrence relation in these matrices.
Using such a relationship, a tree based algorithm (t-SMatPI) is developed, which is shown to be much faster than straightforward computation of the kernel matrices based on their definitions.
This algorithm bypasses the step to compute the SMatPI matrices by other path integral methods and provides more understanding of the SMatPI matrices themselves.
Meanwhile, it keeps the memory cost and computational cost low.
Numerical experiments show that the t-SMatPI algorithm gives exactly the same result as i-QuAPI and SMatPI.
In spite of this, our method may indicate some new properties of open quantum systems, and has the potential to be generalized to higher-order numerical schemes.

\end{abstract}

\section{Introduction}
The system-bath dynamics, which successfully models quantum dissipation and quantum decoherence,
 plays an important role in accurate quantum simulation when the environment plays an non-negligible role.
Most simulations of the system-bath dynamics are based on path integrals.
Due to the large number of paths, a natural approach is the Monte Carlo methods.
For example, the diagrammatic quantum Monte Carlo (dQMC) method applies diagrams
 to intuitively represent the coupling between the system and bath \cite{prokof1998polaron,werner2009diagrammatic}.
For real-time simulations, Monte Carlo based methods suffer from numerical sign problem 
\cite{loh1990sign,cai2022numerical}.
Different techniques, such as application of bold lines \cite{prokof2007bold,chen2017inchworm1,cai2020inchworm}, inclusion-exclusion principle \cite{boag2018inclusion,yang2021inclusion}, combination of thin lines and bold lines \cite{cai2023bold}, are developed recently to relieve the numerical sign problem or accelerate the computation.

Another idea to simulate the system-bath dynamics is to approximate the dynamics by ignoring the long memory effects so that the system depends only on a finite time in the history.
This type of approaches includes the classical
 iterative quasi-adiabatic propagator path integral (i-QuAPI) \cite{makri1992improved,makri1995tensor1,makri1995tensor2}, which is efficient when the decay of the memory kernel is fast.
However, the i-QuAPI method still has to record the contribution of all the paths within the memory length, leading to prohibitively large memory cost when the time nonlocality is long.
Some improvements have been developed in the past decades to reduce the memory cost,
 among which the blip-summed decomposition 
 \cite{makri2014blip,makri2016blip} reduces the memory cost
 by ignoring the paths with small contributions, and
 the differential equation based path integral (DEBPI) studies the continuous form of i-QuAPI and formulates a differential equation system \cite{wang2022differential}.
Recently, the small matrix decomposition of the path integral (SMatPI)
 \cite{makri2020small1,makri2020small2,makri2021small1,makri2021small2}
 successfully overcomes the exponential scaling of the memory cost
 while preserving the accuracy of i-QuAPI.
In this approach, the memory cost scales only linearly with the memory length, leading to a very efficient method for the system-bath dynamics.

The SMatPI method can also be considered as a discrete form of the
 Nakajima-Zwanzig generalized quantum master equation (GQME) \cite{nakajima1958quantum,zwanzig1960ensemble,shi2003new,zhang2006nonequilibrium},
 an integro-differential equation
 with a memory kernel describing the non-Markovian effects.
Formally, the SMatPI method is similar to the transfer tensor method (TTM) 
\cite{cerrillo2014nonmarkovian,buser2017initial,chen2020non}, which also has a time-convolution form.
When the time step $\dt$ tends to zero,
 the SMatPI and the TTM both converge to the GQME \cite{makri2020small1},
 while the special structure of the SMatPI allows 
 it to converge more effectively than the TTM.
Both methods allow extension to more complicated models such as chain models \cite{makri2021small4,wang2023real}.

In this paper, we analyze the structure of the SMatPI matrices and propose a new algorithm for the evaluation of SMatPI matrices directly based on their definitions.
It keeps the memory cost and computational cost low.
The spin-boson model and the SMatPI method are introduced in \cref{sec_spin_boson_SMatPI} and \cref{sec_SMatPI}, respectively. 
In \cref{sec_computational_complexity}, we analyze the structure of SMatPI by combining it to Catalan numbers.
Some recurrence relations of the SMatPI matrices are discovered in \cref{sec_fast_algorithm}, which is then turned into a tree-based algorithm in \cref{sec_implementation}.
Some numerial experiments are carried out in \cref{sec_numerical_results} showing the accuracy and efficiency of our algorithm. Lastly, in \cref{sec_conclusion}, we will summarize our work and discuss how our method may inspire some new theoretical or numerical findings in our future works. 


 
\section{Spin-boson model}
\label{sec_spin_boson_SMatPI}
In a quantum system,
 the density matrix $\rho(t)$ satisfies the von Neumann equation
\begin{equation*}
    \ii\dv{\rho(t)}{t} = H\rho(t) - \rho(t) H
\end{equation*}
where $\ii$ is the imaginary unit and $H$ is the Hamiltonian.
In a general open quantum system, the Hamiltonian is defined on the tensor product space $\mathcal{H}_s \otimes \mathcal{H}_b$, where $\mathcal{H}_s$ and $\mathcal{H}_b$ represent the Hilbert spaces for the system and the bath, respectively.

A fundamental example of open quantum systems
 is the spin-boson model where the Hilbert spaces are
\begin{equation*}
    \mathcal{H}_s = \SPAN\{\ket{-1},\ket{+1}\}, \quad
    \mathcal{H}_b = \bigotimes_{j=1}^L (\mathcal{L}^2(\mathbb{R}^3))
\end{equation*}
with $\mathcal{L}^2(\mathbb{R}^3)$ being the $\mathcal{L}^2$ space
over $\mathbb{R}^3$ and $L$ being the total number of harmonic oscillators in the bath.
In the spin-boson model,
 the Hamiltonian can always be written as the sum of three parts:
\begin{equation*}
    H = \left(\epsilon\sigmaz + \Delta\sigmax\right) \otimes \id_b + \id_s \otimes \left(\sum_{j}\frac{1}{2}\left(
    \hat{p}_j^2 + \omega_j^2\hat{q}_j^2
    \right)\right)
    + \sigmaz \otimes \left(\sum_j c_j\hat{q}_j\right)
\end{equation*}
where $\id_s$ and $\id_b$ are the identity operators on $\mathcal{H}_s$ and $\mathcal{H}_b$ respectively,
$\sigmax,\sigmaz$ are Pauli matrices,
and $\hat{p}_j,\hat{q}_j$ are momentum operator and position operator
 of the $j$th harmonic oscillator, respectively.
$\epsilon$ represents the energy difference between two spin states
and $\Delta$ is the frequency of the spin flipping.
The parameter $\omega_j$ is the frequency of the $j$th harmonic oscillator while $c_j$ is the coupling intensity between the spin and the $j$th harmonic oscillator.

The path integral with the influence functional \cite{feynman1963theory} suggests that
 the density matrix of the system $\rho_s$ has the following representation \cite{makri1995numerical,makri1998quantum}:
\begin{equation*}
\begin{split}
    \rho_s(N\dt) = \sum_{\sigma_0^\pm,\dots,\sigma_N^\pm}
    &\ket{\sigma_N^+} \mel{\sigma_N^+}{\e^{-\ii H_0\dt}}{\sigma_{N-1}^+} \dots
    \mel{\sigma_1^+}{\e^{-\ii H_0\dt}}{\sigma_0^+}
    \mel{\sigma_0^+}{\rho_s(0)}{\sigma_0^-} \\
    &\times\mel{\sigma_0^-}{\e^{\ii H_0\dt}}{\sigma_1^-}
    \dots
    \mel{\sigma_{N-1}^-}{\e^{\ii H_0 \dt}}{\sigma_N^-}
    \bra{\sigma_N^-}
    F(\boldsymbol{\sigma}^+,\boldsymbol{\sigma}^-)
\end{split}
\end{equation*}
where $\boldsymbol{\sigma}^\pm=(\sigma_0^\pm,\dots,\sigma_N^\pm)$
 is the ``path'';
 $\dt$ is the time step;
 $\displaystyle H_0 = \epsilon\hat{\sigma}_z + \Delta \hat{\sigma}_x - \sum_j \frac{c_j \hat{\sigma}_z}{2\omega_j^2}$ is called reference Hamiltonian \cite{makri1998quantum};
 $F$ is the influence functional with the following form \cite{feynman1963theory,makri1998quantum}:
\begin{equation*}
    F(\boldsymbol{\sigma}^+,\boldsymbol{\sigma}^-)
    = \prod_{j_1=0}^N
    \prod_{j_2 = 0}^{j_1}
    \exp\left(-\left(\sigma_{j_1}^+ - \sigma_{j_1}^-\right)
    \left(\eta_{j_1,j_2} \sigma_{j_2}^+ - \overline{\eta_{j_1,j_2}} \sigma_{j_2}^-\right)\right).
\end{equation*}
Here $\overline{\eta}$ is the complex conjugate of $\eta$.
The complex function $\eta$,
relying on the spectral density $J(\omega)$ and thermodynamic parameter $\beta$,
takes different forms for different $j_1$ and $j_2$ (see \cite[Section \MakeUppercase{\romannumeral 3}]{makri1995tensor1} for more details).

\section{Small matrix decomposition of the path integral}
\label{sec_SMatPI}
In \cite{makri2020small1},
 the small matrix decomposition of the path integral expression (SMatPI) is developed.
It solves the problem of exponentially growing memory cost in many path integral based methods 
 like the well-known i-QuAPI method \cite{makri1995tensor1,makri1995tensor2}.
In this section, we introduce the SMatPI method and carry out some analysis on its computational cost.

The author of \cite{makri2020small1} introduces a reduced propagator
\begin{equation}
\label{eq_UN0_original}
    \U{N}{0}
    = \sum_{\sigma_{N-1}^\pm=\pm1}
    \dots
    \sum_{\sigma_{1}^\pm=\pm1}
    \left[\G{N}{N-1}\dots\G{1}{0}
    \prod_{j_1 = 0}^{N} \prod_{j_2=0}^{j_1}
    \F{j_1}{j_2}
    \right]
\end{equation}
where
\begin{equation*}
    \G{k+1}{k} = \mel{\sigma_{k+1}^+}{\e^{-\ii H_0\dt}}{\sigma_{k}^+}
    \mel{\sigma_{k}^-}{\e^{\ii H_0 \dt}}{\sigma_{k+1}^-}
\end{equation*}
and
\begin{equation*}
    \F{j_1}{j_2} = \exp\left(
    -\left(\sigma_{j_1}^+ - \sigma_{j_1}^-\right)
    \left(\eta_{j_1,j_2} \sigma_{j_2}^+ - \overline{\eta_{j_1,j_2}} \sigma_{j_2}^-\right)
    \right).
\end{equation*}
The reduced density matrix can be then
 computed by the propagator $\U{N}{0}$:
\begin{equation}
\label{eq_density}
    \rho_{\sigma_N^{\pm}}(N\dt)
    = \sum_{\sigma_0^\pm=\pm1}
    \U{N}{0} \rho_{\sigma_0^\pm}.
\end{equation}

\begin{figure}
    \centering
    \begin{tikzpicture}[baseline={(current bounding box.center)}]
	\draw (0,0) -- (4,4);
	\draw (0,0) -- (4,0);
	\draw (4,0) -- (4,4);
	\draw (1.3333,0) -- (1.3333,1.3333);
	\draw (1.3333,1.3333) -- (4,1.3333);
	\node at (8/9,4/9) {${(0,0)}$};
    \node at (8/3,2/3) {${(1,0)}$};
    \node at (28/9,20/9) {${(1,1)}$};
\end{tikzpicture}
    \caption{Diagrammatic representation of $\U{1}{0}$.
    Each pair of indices represents an $F$ in the multiplication.
    If the difference of two indices is one, an extra $G$ should be also multiplied. This diagram represents $\F{0}{0}\F{1}{0}\G{1}{0}\F{1}{1}$.
    Note that the different shapes come from the different expressions of $\eta$ in \cite{makri1995tensor1}.}
    \label{fig_U10}
\end{figure}
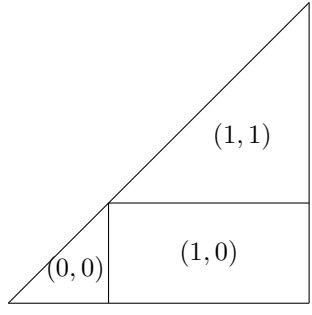

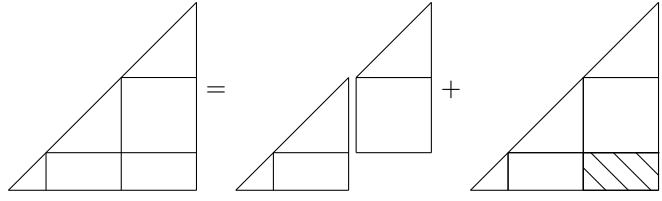
\begin{figure}
    \centering
    \begin{tikzpicture}[baseline={(current bounding box.center)}]
	\draw (0,0) -- (2.5,2.5);
	\draw (0,0) -- (2.5,0);
	\draw (2.5,0) -- (2.5,2.5);
	\draw (0.5,0) -- (0.5,0.5);
	\draw (0.5,0.5) -- (2.5,0.5);
	\draw (1.5,0) -- (1.5,1.5);
	\draw (1.5,1.5) -- (2.5,1.5);
\end{tikzpicture}
=
\begin{tikzpicture}[baseline={(current bounding box.center)}]
	\draw (0,0) -- (1.5,1.5);
	\draw (0,0) -- (1.5,0);
	\draw (1.5,0) -- (1.5,1.5);
	\draw (0.5,0) -- (0.5,0.5);
	\draw (0.5,0.5) -- (1.5,0.5);
	\draw (1.6,0.5) -- (1.6,1.5);
	\draw (1.6,1.5) -- (2.6,1.5);
	\draw (2.6,1.5) -- (2.6,0.5);
	\draw (2.6,0.5) -- (1.6,0.5);
	\draw (1.6,1.5) -- (2.6,2.5);
	\draw (2.6,2.5) -- (2.6,1.5);
	\draw (2.6,1.5) -- (1.6,1.5);
\end{tikzpicture}
+
\begin{tikzpicture}[baseline={(current bounding box.center)}]
	\draw (0,0) -- (0.5,0.5);
	\draw (0.5,0.5) -- (0.5,0);
	\draw (0.5,0) -- (0,0);
	\draw (0.5,0) -- (0.5,0.5);
	\draw (0.5,0.5) -- (1.5,0.5);
	\draw (1.5,0.5) -- (1.5,0);
	\draw (1.5,0) -- (0.5,0);
	\draw (0.5,0.5) -- (1.5,1.5);
	\draw (1.5,1.5) -- (1.5,0.5);
	\draw (1.5,0.5) -- (0.5,0.5);
	\draw (1.5,0) -- (1.5,0.5);
	\draw (1.5,0.5) -- (2.5,0.5);
	\draw (2.5,0.5) -- (2.5,0);
	\draw (2.5,0) -- (1.5,0);
	\draw (1.5,0) -- (1.5,0);
	\draw (1.8,0) -- (1.5,0.3);
	\draw (2.1,0) -- (1.6,0.5);
	\draw (2.4,0) -- (1.9,0.5);
	\draw (2.5,0.2) -- (2.2,0.5);
	\draw (2.5,0.5) -- (2.5,0.5);
	\draw (1.5,0) -- (1.5,0.5);
	\draw (1.5,0.5) -- (2.5,0.5);
	\draw (2.5,0.5) -- (2.5,0);
	\draw (2.5,0) -- (1.5,0);
	\draw (1.5,0.5) -- (1.5,1.5);
	\draw (1.5,1.5) -- (2.5,1.5);
	\draw (2.5,1.5) -- (2.5,0.5);
	\draw (2.5,0.5) -- (1.5,0.5);
	\draw (1.5,1.5) -- (2.5,2.5);
	\draw (2.5,2.5) -- (2.5,1.5);
	\draw (2.5,1.5) -- (1.5,1.5);
\end{tikzpicture}
    \caption{Diagrammatic representation of \cref{eq_U20}. The first figure on the right hand side represents the term $\M{2}{1}\U{1}{0}$
    while the second term represents $\M{2}{0}$.
    The shaded rectangle in second term represents the $\F{2}{0}-1$ in \cref{eq_U20}, which turns out to be in $\M{2}{0}$.
    Note that \cref{eq_U20} also takes summation over $\sigma_1^\pm$ for both sides.
    For simplicity, the indices are omitted in the figures.}
    \label{fig_U20}
\end{figure}

The reduced propagators are computed by recurrence relations. 
To begin with, we define a few matrices $\mathbf{U}^{(1,0)}$, $\mathbf{M}^{(1,0)}$ and $\mathbf{A}^{(1,0)}$ elementwisely by 
\begin{equation*}
    \U{1}{0} \coloneqq \M{1}{0} \coloneqq
    \A{1}{0} \coloneqq
    \G{1}{0}\F{1}{1}\F{1}{0}\F{0}{0}.
\end{equation*}
For better illustration, $\U{1}{0}$ is also represented diagrammatically by \cref{fig_U10}.
The next propagator $\mathbf{U}^{(2,0)}$ is then computed by
\begin{equation*}
\begin{split}
    \U{2}{0} &= \sum_{\sigma_1^\pm = \pm 1}
    \G{2}{1}\G{1}{0}
    \F{2}{0}\F{2}{1}\F{2}{2}
    \F{1}{0}\F{1}{1}\F{0}{0} \\
    &= \F{2}{0} \sum_{\sigma_1^\pm=\pm 1}
    \M{2}{1} \U{1}{0}
\end{split}
\end{equation*}
where 
\begin{equation*}
    \M{k+1}{k} \coloneqq \A{k+1}{k} \coloneqq \F{k+1}{k}\F{k+1}{k+1}\G{k+1}{k}
\end{equation*}
for $k=1,2,\dots$.
The SMatPI then splits $\F{2}{0}$ to be the sum of $1$ and $(\F{2}{0}-1)$:
\begin{equation}
\label{eq_U20}
\begin{split}
    \U{2}{0} 
    &= \sum_{\sigma_1^\pm=\pm 1} 
    \M{2}{1}\U{1}{0}
    + \sum_{\sigma_1^\pm=\pm1}
    (\F{2}{0}-1) \A{2}{1} \A{1}{0} \\
    &\eqqcolon \sum_{\sigma_1^\pm=\pm 1} 
    \M{2}{1}\U{1}{0}
    + \M{2}{0}.
\end{split}
\end{equation}
Similar to $\U{1}{0}$,
 the diagrammatic representation of \cref{eq_U20} is in \cref{fig_U20}
 where the shaded box represents the term $(\F{2}{0}-1)$.
We can also write \cref{eq_U20} in the matrix form:
\begin{equation*}
    \mathbf{U}^{(2,0)} = \mathbf{M}^{(2,1)}\mathbf{U}^{(1,0)} + \mathbf{M}^{(2,0)}.
\end{equation*}
Following this procedure, the elements of $\mathbf{U}^{(3,0)}$ can be represented by
\begin{equation}
    \label{eq_U30}
    \U{3}{0} = \sum_{\sigma_1^\pm = \pm 1}\M{3}{1}\U{1}{0}
    + \sum_{\sigma_2^\pm = \pm 1}\M{3}{2}\U{2}{0}
    + \M{3}{0}.
\end{equation}
where $\M{3}{1}$ is given by
\begin{equation*}
    \M{k+2}{k} \coloneqq (\F{k+2}{k}-1) 
    \sum_{\sigma_{k+1}^\pm = \pm 1}
    \A{k+2}{k+1}\A{k+1}{k}, \qquad k \geqslant 1,
\end{equation*}
and
\begin{equation}
\label{eq_M30}
\begin{split}
    \M{3}{0} \coloneqq \sum_{\sigma_2^\pm = \pm 1}\sum_{\sigma_1^\pm = \pm 1}
    &\left[
    (\F{3}{0}-1)\F{4}{2}\F{3}{1}\A{3}{2}\A{2}{1}\A{1}{0}\right. \\
    &\left.+ 
    (\F{3}{1}-1)(\F{2}{0}-1)\A{3}{2}\A{2}{1}\A{1}{0}
    \right].
\end{split}
\end{equation}
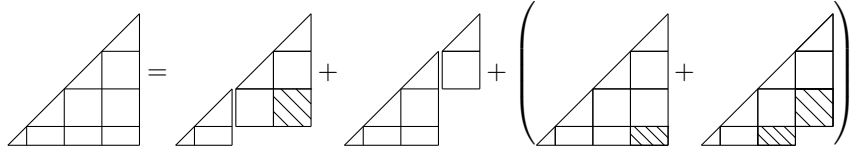
\begin{figure}
    \centering
    \begin{equation*}
\begin{tikzpicture}[baseline={(current bounding box.center)},scale=0.5]
	\draw (0,0) -- (3.5,3.5);
	\draw (0,0) -- (3.5,0);
	\draw (3.5,0) -- (3.5,3.5);
	\draw (0.5,0) -- (0.5,0.5);
	\draw (0.5,0.5) -- (3.5,0.5);
	\draw (1.5,0) -- (1.5,1.5);
	\draw (1.5,1.5) -- (3.5,1.5);
	\draw (2.5,0) -- (2.5,2.5);
	\draw (2.5,2.5) -- (3.5,2.5);
\end{tikzpicture}
=
\begin{tikzpicture}[baseline={(current bounding box.center)},scale=0.5]
	\draw (0,0) -- (1.5,1.5);
	\draw (0,0) -- (1.5,0);
	\draw (1.5,0) -- (1.5,1.5);
	\draw (0.5,0) -- (0.5,0.5);
	\draw (0.5,0.5) -- (1.5,0.5);
	\draw (1.6,0.5) -- (1.6,1.5);
	\draw (1.6,1.5) -- (2.6,1.5);
	\draw (2.6,1.5) -- (2.6,0.5);
	\draw (2.6,0.5) -- (1.6,0.5);
	\draw (1.6,1.5) -- (2.6,2.5);
	\draw (2.6,2.5) -- (2.6,1.5);
	\draw (2.6,1.5) -- (1.6,1.5);
	\draw (2.6,0.5) -- (2.6,1.5);
	\draw (2.6,1.5) -- (3.6,1.5);
	\draw (3.6,1.5) -- (3.6,0.5);
	\draw (3.6,0.5) -- (2.6,0.5);
	\draw (2.6,0.5) -- (2.6,0.5);
	\draw (3,0.5) -- (2.6,0.9);
	\draw (3.4,0.5) -- (2.6,1.3);
	\draw (3.6,0.7) -- (2.8,1.5);
	\draw (3.6,1.1) -- (3.2,1.5);
	\draw (3.6,1.5) -- (3.6,1.5);
	\draw (2.6,1.5) -- (2.6,2.5);
	\draw (2.6,2.5) -- (3.6,2.5);
	\draw (3.6,2.5) -- (3.6,1.5);
	\draw (3.6,1.5) -- (2.6,1.5);
	\draw (2.6,2.5) -- (3.6,3.5);
	\draw (3.6,3.5) -- (3.6,2.5);
	\draw (3.6,2.5) -- (2.6,2.5);
\end{tikzpicture}
+
\begin{tikzpicture}[baseline={(current bounding box.center)},scale=0.5]
	\draw (0,0) -- (2.5,2.5);
	\draw (0,0) -- (2.5,0);
	\draw (2.5,0) -- (2.5,2.5);
	\draw (0.5,0) -- (0.5,0.5);
	\draw (0.5,0.5) -- (2.5,0.5);
	\draw (1.5,0) -- (1.5,1.5);
	\draw (1.5,1.5) -- (2.5,1.5);
	\draw (2.6,1.5) -- (2.6,2.5);
	\draw (2.6,2.5) -- (3.6,2.5);
	\draw (3.6,2.5) -- (3.6,1.5);
	\draw (3.6,1.5) -- (2.6,1.5);
	\draw (2.6,2.5) -- (3.6,3.5);
	\draw (3.6,3.5) -- (3.6,2.5);
	\draw (3.6,2.5) -- (2.6,2.5);
\end{tikzpicture}
+
\left(
\begin{tikzpicture}[baseline={(current bounding box.center)},scale=0.5]
	\draw (0,0) -- (0.5,0.5);
	\draw (0.5,0.5) -- (0.5,0);
	\draw (0.5,0) -- (0,0);
	\draw (0.5,0) -- (0.5,0.5);
	\draw (0.5,0.5) -- (1.5,0.5);
	\draw (1.5,0.5) -- (1.5,0);
	\draw (1.5,0) -- (0.5,0);
	\draw (0.5,0.5) -- (1.5,1.5);
	\draw (1.5,1.5) -- (1.5,0.5);
	\draw (1.5,0.5) -- (0.5,0.5);
	\draw (1.5,0) -- (1.5,0.5);
	\draw (1.5,0.5) -- (2.5,0.5);
	\draw (2.5,0.5) -- (2.5,0);
	\draw (2.5,0) -- (1.5,0);
	\draw (1.5,0.5) -- (1.5,1.5);
	\draw (1.5,1.5) -- (2.5,1.5);
	\draw (2.5,1.5) -- (2.5,0.5);
	\draw (2.5,0.5) -- (1.5,0.5);
	\draw (1.5,1.5) -- (2.5,2.5);
	\draw (2.5,2.5) -- (2.5,1.5);
	\draw (2.5,1.5) -- (1.5,1.5);
	\draw (2.5,0) -- (2.5,0.5);
	\draw (2.5,0.5) -- (3.5,0.5);
	\draw (3.5,0.5) -- (3.5,0);
	\draw (3.5,0) -- (2.5,0);
	\draw (2.5,0) -- (2.5,0);
	\draw (2.8,0) -- (2.5,0.3);
	\draw (3.1,0) -- (2.6,0.5);
	\draw (3.4,0) -- (2.9,0.5);
	\draw (3.5,0.2) -- (3.2,0.5);
	\draw (3.5,0.5) -- (3.5,0.5);
	\draw (2.5,0) -- (2.5,0.5);
	\draw (2.5,0.5) -- (3.5,0.5);
	\draw (3.5,0.5) -- (3.5,0);
	\draw (3.5,0) -- (2.5,0);
	\draw (2.5,0.5) -- (2.5,1.5);
	\draw (2.5,1.5) -- (3.5,1.5);
	\draw (3.5,1.5) -- (3.5,0.5);
	\draw (3.5,0.5) -- (2.5,0.5);
	\draw (2.5,1.5) -- (2.5,2.5);
	\draw (2.5,2.5) -- (3.5,2.5);
	\draw (3.5,2.5) -- (3.5,1.5);
	\draw (3.5,1.5) -- (2.5,1.5);
	\draw (2.5,2.5) -- (3.5,3.5);
	\draw (3.5,3.5) -- (3.5,2.5);
	\draw (3.5,2.5) -- (2.5,2.5);
\end{tikzpicture}
+
\begin{tikzpicture}[baseline={(current bounding box.center)},scale=0.5]
	\draw (0,0) -- (0.5,0.5);
	\draw (0.5,0.5) -- (0.5,0);
	\draw (0.5,0) -- (0,0);
	\draw (0.5,0) -- (0.5,0.5);
	\draw (0.5,0.5) -- (1.5,0.5);
	\draw (1.5,0.5) -- (1.5,0);
	\draw (1.5,0) -- (0.5,0);
	\draw (0.5,0.5) -- (1.5,1.5);
	\draw (1.5,1.5) -- (1.5,0.5);
	\draw (1.5,0.5) -- (0.5,0.5);
	\draw (1.5,0) -- (1.5,0.5);
	\draw (1.5,0.5) -- (2.5,0.5);
	\draw (2.5,0.5) -- (2.5,0);
	\draw (2.5,0) -- (1.5,0);
	\draw (1.5,0) -- (1.5,0);
	\draw (1.8,0) -- (1.5,0.3);
	\draw (2.1,0) -- (1.6,0.5);
	\draw (2.4,0) -- (1.9,0.5);
	\draw (2.5,0.2) -- (2.2,0.5);
	\draw (2.5,0.5) -- (2.5,0.5);
	\draw (1.5,0) -- (1.5,0.5);
	\draw (1.5,0.5) -- (2.5,0.5);
	\draw (2.5,0.5) -- (2.5,0);
	\draw (2.5,0) -- (1.5,0);
	\draw (1.5,0.5) -- (1.5,1.5);
	\draw (1.5,1.5) -- (2.5,1.5);
	\draw (2.5,1.5) -- (2.5,0.5);
	\draw (2.5,0.5) -- (1.5,0.5);
	\draw (1.5,1.5) -- (2.5,2.5);
	\draw (2.5,2.5) -- (2.5,1.5);
	\draw (2.5,1.5) -- (1.5,1.5);
	\draw (2.5,0.5) -- (2.5,1.5);
	\draw (2.5,1.5) -- (3.5,1.5);
	\draw (3.5,1.5) -- (3.5,0.5);
	\draw (3.5,0.5) -- (2.5,0.5);
	\draw (2.5,1.5) -- (2.5,2.5);
	\draw (2.5,2.5) -- (3.5,2.5);
	\draw (3.5,2.5) -- (3.5,1.5);
	\draw (3.5,1.5) -- (2.5,1.5);
	\draw (2.5,2.5) -- (3.5,3.5);
	\draw (3.5,3.5) -- (3.5,2.5);
	\draw (3.5,2.5) -- (2.5,2.5);
	\draw (2.5,0.5) -- (2.5,1.5);
	\draw (2.5,1.5) -- (3.5,1.5);
	\draw (3.5,1.5) -- (3.5,0.5);
	\draw (3.5,0.5) -- (2.5,0.5);
	\draw (2.5,0.5) -- (2.5,0.5);
	\draw (2.9,0.5) -- (2.5,0.9);
	\draw (3.3,0.5) -- (2.5,1.3);
	\draw (3.5,0.7) -- (2.7,1.5);
	\draw (3.5,1.1) -- (3.1,1.5);
	\draw (3.5,1.5) -- (3.5,1.5);
	\draw (2.5,0.5) -- (2.5,1.5);
	\draw (2.5,1.5) -- (3.5,1.5);
	\draw (3.5,1.5) -- (3.5,0.5);
	\draw (3.5,0.5) -- (2.5,0.5);
	\draw (2.5,1.5) -- (2.5,2.5);
	\draw (2.5,2.5) -- (3.5,2.5);
	\draw (3.5,2.5) -- (3.5,1.5);
	\draw (3.5,1.5) -- (2.5,1.5);
	\draw (2.5,2.5) -- (3.5,3.5);
	\draw (3.5,3.5) -- (3.5,2.5);
	\draw (3.5,2.5) -- (2.5,2.5);
\end{tikzpicture}
\right)
\end{equation*}
    \caption{Diagrammatic representation of \cref{eq_U30}.
    The first figure on the right hand side represents the term $\M{3}{1}\U{1}{0}$;
    the second figure is the term $\M{3}{2}\U{2}{1}$;
    the last two figures on the right hand sides together represent $\M{3}{0}$,
    corresponding to the two terms in \cref{eq_M30}.}
    \label{fig_U30}
\end{figure}%
The corresponding matrix form is:
\begin{equation*}
    \mathbf{U}^{(3,0)} = \mathbf{M}^{(3,2)}\mathbf{U}^{(2,0)}
    + \mathbf{M}^{(3,1)}\mathbf{U}^{(1,0)}
    + \mathbf{M}^{(3,0)}.
\end{equation*}
\Cref{fig_U30} shows the diagrammatic representation for \cref{eq_U30}.
The same idea can be further extended to $\mathbf{U}^{(r,0)}$ for arbitrary
$r\geqslant 1$ and we then have
\begin{equation}
\label{Ur0}
    \mathbf{U}^{(r,0)} = \sum_{m=1}^{r-1} \mathbf{M}^{(r,m)}\mathbf{U}^{(m,0)}
    + \mathbf{M}^{(r,0)}.
\end{equation}
In addition, the matrix $\mathbf{M}^{(r,m)}$ has the following shift invariance
\begin{equation}
\label{shift_invariance_M}
    \mathbf{M}^{(k,0)} \not= \mathbf{M}^{(k+1,1)}
    = \mathbf{M}^{(k+2,2)} = \mathbf{M}^{(k+3,3)} = \dots.
\end{equation}
based on the intrinsic shift invariances of $\G{k+1}{k}$ and $\F{j_1}{j_2}$.
These invariance are also applied in several other research \cite{cai2022fast,cai2023bold}.
Although $\mathbf{M}^{(k,0)}\not=\mathbf{M}^{(k+1,1)}$,
 they have very similar forms.
Compare to the expression of $\M{k}{0}$,
 all the indices in $\M{k+1}{1}$ is increased by 1.
The difference of $\mathbf{M}^{(k,0)}$ and $\mathbf{M}^{(k+1,1)}$ is because $A_{s_1^\pm,s_2^\pm}^{(1,0)}\not=A_{s_1^\pm,s_2^\pm}^{(2,1)}$ for $s_1^\pm,s_2^\pm\in\{-1,+1\}$.

The computation can be further simplified by truncating 
 the memory length $\Delta k_{\mathrm{max}}$
 and the temporal entanglement length $r_{\mathrm{max}}$.
For simplicity, we choose $\Delta k_{\mathrm{max}} = r_{\mathrm{max}}$
 and we use $\Delta k$ to represent both (see \cite[Section \MakeUppercase{\romannumeral 4}]{makri2020small1} for more details).
With the truncations, $\mathbf{U}^{(r,0)}$ can be computed based on
\begin{equation}
\label{Ur0_truncated}
    \mathbf{U}^{(r,0)} = \sum_{m=1}^{\Delta k} \mathbf{M}^{(r,r-m)} \mathbf{U}^{(r-m,0)}
    = \sum_{m=1}^{\Delta k} 
    \mathbf{M}^{(m+1,1)} \mathbf{U}^{(r-m,0)}
\end{equation}
when $r = \Delta k + 1,\Delta k + 2,\dots$.

After precomputing $\mathbf{M}^{(k,0)}$ 
 and $\mathbf{M}^{(k+1,1)}$ for $k=1,\dots,\Delta k$,
 the reduced propagator matrix $\mathbf{U}^{(r,0)}$
 for any $r\geqslant 1$
 by either \cref{Ur0} or \cref{Ur0_truncated}.
Therefore, compared with the QuAPI, the SMatPI is free from the exponentially growth of memory cost.
Instead, the required memory cost in the SMatPI
 grows linearly
 with respect to the memory length $\dk$,
 making it possible to choose longer memory length 
 or smaller time step $\dt$.

Although the previous discussion gives
 the definition of SMatPI matrices $\mathbf{M}^{(k,0)}$, $\mathbf{M}^{(k+1,1)}$,
 the practical computation of these matrices are not based directly on their expressions.
In \cite{makri2020small1},
 the author proposes a method for the evaluation of SMatPI matrices.
This method requires precomputation of $\mathbf{U}^{(r,0)}$ for $r=1,2,\dots,\dk$ by QuAPI or other path integral methods.
$\mathbf{M}^{(N,N-r)}=\mathbf{M}^{(r+1,1)}$ can be obtained from the expression for $\mathbf{M}^{(r,0)}$ if the influence functional coefficients $\eta_{k,0}$ are replaced by $\eta_{k+1,1}$.
By \cref{eq_U20}, $\mathbf{M}^{(2,0)}$ can be computed by subtraction given three matrices $\mathbf{M}^{(2,1)},\mathbf{U}^{(1,0)}$ and $\mathbf{U}^{(2,0)}$.
Matrix $\mathbf{M}^{(3,1)}$ can then be obtained and $\mathbf{M}^{(3,0)}$ can be computed by \cref{eq_U30}.
One can repeat this process until all SMatPI matrices $\mathbf{M}^{(k+1,1)}$ for $k=1,\dots,\dk$ are obtained.

Since the computation cost of $\mathbf{M}^{(k,0)}$ from \cref{Ur0} is clearly of linear order of $\dk$,
the main computational cost then comes from the preparation of $\mathbf{U}^{(r,0)}$ for $r=1,\dots,\dk$.
Since we do not apply any truncation in the computation of $\mathbf{U}^{(r,0)}$ for $r=1,\dots, \dk$,
 the algorithm applied here is full memory QuAPI or full memory blip-summed \cite{kundu2023pathsum}.
Let us just take QuAPI as an example
and focus on the computation of $\mathbf{U}^{(k,0)}$ defined in \cref{eq_UN0_original}.
Computing $\U{k}{0}$ directly by \cref{eq_UN0_original} involving the summation over $\sigma_1^\pm,\dots,\sigma_{k-1}^\pm$.
For a two-level system, the cost for the summand is of order $O(k^2)$ from the product of $\F{j_1}{j_2}$.
Therefore, a naïve method to compute $\mathbf{U}^{(k,0)}$ for $k=1,\dots,\dk$ is of order
\begin{equation}
    \sum_{k=1}^{\dk}k^2 2^{2k+2} = O(\dk^2 4^{\dk}).
\end{equation}
By this method, 
the computations of different paths are totally independent 
so that the computation can be easily parallelized based on different path.
This idea is applied in the \textsc{PathSum} package introduced in \cite{kundu2023pathsum} for parallel QuAPI computation \cite{makri1995numerical,makri2014blip,makri2020small2}.
\begin{remark}
    Here, we would like to remark that the idea do not require storage of values $\U{k}{0}$ for all paths.
    However, the value $\U{k}{0}$ for a specific path can be discarded after adding it into the final matrix $\mathbf{U}^{(k,0)}$.
    Therefore, this method does not require high memory cost in the step of computing $\mathbf{U}^{(k,0)}$.
\end{remark}

The package \textsc{PathSum} also provides a single thread algorithm for the computation of QuAPI.
In this implementation, the paths are extended step by step.
Since \cref{eq_UN0_original} is defined as a summation over all possible paths, we can simply rewrite it as
\begin{equation}
    \U{k}{0} = \sum_{\sigma_{k-1}^\pm=\pm 1}
    \dots \sum_{\sigma_1^\pm=\pm 1}
    \Ucal{k}{0}
\end{equation}
where $\Ucal{k}{0}$ only relies on the zeroth to the $k$th components of a fixed path $\boldsymbol{\sigma}^\pm = (\sigma_0^\pm,\dots,\sigma_{\dk}^\pm)$
and it is clear that $\Ucal{k}{0}$ has the following definition
\begin{equation}
    \Ucal{k}{0} = \G{k}{k-1}\dots\G{1}{0}
    \prod_{j_1 = 0}^{k} \prod_{j_2=0}^{j_1}
    \F{j_1}{j_2}.
\end{equation}
By extending the path $\sigma(0:k)$ to $\sigma(0:k+1)$,
 we have the following recurrence relation:
\begin{equation}
\begin{split}
    \Ucal{k+1}{0} &= \G{k+1}{k} \G{k}{k-1}\dots\G{1}{0}
    \prod_{j_1 = 0}^{k+1} \prod_{j_2=0}^{j_1}
    \F{j_1}{j_2} \\
    &= \left(
    \G{k}{k-1}\dots\G{1}{0} \prod_{j_1 = 0}^{k} \prod_{j_2=0}^{j_1}
    \right) \G{k+1}{k} \prod_{j_2=0}^{k+1} \F{k+1}{j_2} 
    = \Ucal{k}{0} \G{k+1}{k} \prod_{j_2=0}^{k+1} \F{k+1}{j_2}.
\end{split}
\end{equation}
Based on this recurrence relation,
 if $\Ucal{k}{0}$ is known,
 the computational cost of $\Ucal{k+1}{0}$ is of order $O(k)$.
Then the total computational cost of QuAPI based on this recurrence relation is
\begin{equation}
    \sum_{k=1}^{\dk} k 2^{2k+2} = O(\dk 4^{\dk}).
\end{equation}
This recurrence is adopted in the single thread computation of QuAPI in \textsc{PathSum} package.
\begin{remark}
    Here, we would also like to discuss the possibility to parallelize the computation based on this recurrence relation.
    Since the quantities $\Ucal{k+1}{0}$ relies on the shorter path $\Ucal{k}{0}$,
    the computations of $\Ucal{k+1}{0}$ and $\Ucall{k+1}{0}$ rely on different quantities, $\Ucal{k}{0}$ and $\Ucall{k}{0}$, respectively 
    in the case where $\sigma(0\colon k)$ and $\varsigma(0\colon k)$ are not exactly identical.
    The computation of these values can be parallelized by carefully assigning the tasks to each thread although the parallelization is not implemented in \textsc{PathSum}.
    A weakness of this approach is
     the memory cost.
    In order to compute the $\mathbf{U}^{(k+1,0)}$,
     $\Ucal{k+1}{0}$ of all possible paths should be computed,
     and therefore, $\Ucal{k}{0}$ of all possible paths should be stored, leading to the memory cost of order $O(4^{\dk})$ in two-level systems.
\end{remark}

\begin{figure}
    \centering
    \includegraphics[width=0.5\textwidth]{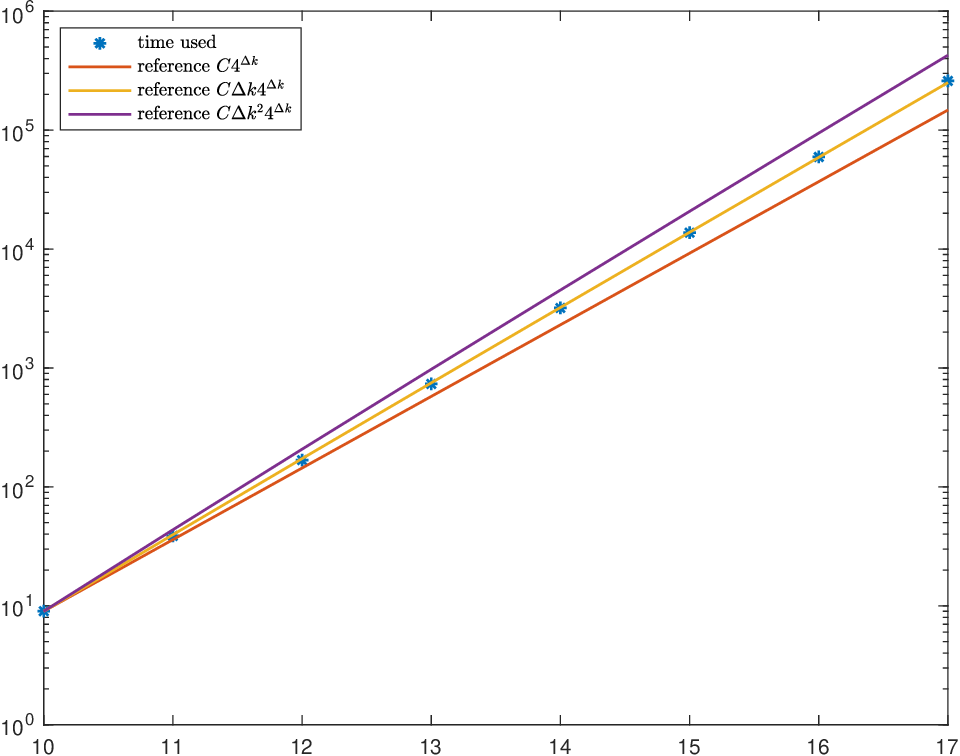}
    \caption{Computational time of SMatPI method by \textsc{PathSum}.}
    \label{fig_SMatPI_running_time}
\end{figure}

To better understand the computational complexity of SMatPI in the \textsc{PathSum} package \cite{kundu2023pathsum,makri2020small1,makri2020small2,makri2021small3}, we carry out numerical tests with different $\dk$ and clock its running time. The result is then shown in \cref{fig_SMatPI_running_time}.
Each experiment is repeated five times and the average time is adopted to plot \cref{fig_SMatPI_running_time}. 
Based on the result, the running time for SMatPI in \textsc{PathSum} grows as the order $O(\dk 4^{\dk})$.

\section{Number of diagrams in SMatPI matrices}
\label{sec_computational_complexity}
In this section, we work on the direct computation cost of SMatPI matrices by their definitions. The structure of their definitions also inspire us to design an alternative method for evaluating the SMatPI matrices in the following sections.

\begin{figure}
    \input{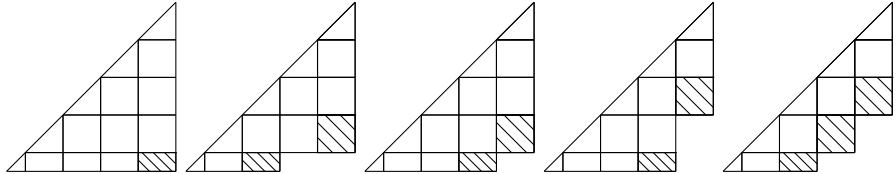}
    \caption{All terms in $\M{4}{0}$.
    There are in total 5 different staircases.}
    \label{fig_M40}
\end{figure}

Since the definitions of $\mathbf{M}^{(k,0)}$ and $\mathbf{M}^{(k+1,1)}$ are very similar,
 we only work on the matrices $\mathbf{M}^{(k,0)}$
 and the analysis for $\mathbf{M}^{(k+1,1)}$ is almost identical.
We first investigate the number of diagrams involved in the definition of $\mathbf{M}^{(k,0)}$ for $k=1,\dots,\dk$.
In $\mathbf{M}^{(3,0)}$, there are two terms for each path as shown in the bracket in \cref{fig_U30} or in \cref{eq_M30}.
In $\mathbf{M}^{(4,0)}$, there are in total five terms (see \cref{fig_M40}) for each path.
When $k = 5$, the number of terms in $\mathbf{M}^{(5,0)}$ increases to 14.
We will show below that the number of terms in $\mathbf{M}^{(k,0)}$ increases exponentially fast with respect to $k$.

\begin{figure}
    \begin{tikzpicture}[scale = 0.8]
  \draw[step=1cm,gray,very thin] (-0.1,-0.1) grid (3.1,3.1);
  \draw [gray,very thin](0,0) -- (3,3);
  \filldraw (0,0) circle (.1cm);
  \filldraw (3,3) circle (.1cm);
  \draw[line width=1.5pt] (0,0) -- (1,0) -- (2,0) -- (3,0) -- (3,1) -- (3,2) -- (3,3);
\end{tikzpicture}
\begin{tikzpicture}[scale = 0.8]
  \draw[step=1cm,gray,very thin] (-0.1,-0.1) grid (3.1,3.1);
  \draw [gray,very thin](0,0) -- (3,3);
  \filldraw (0,0) circle (.1cm);
  \filldraw (3,3) circle (.1cm);
  \draw[line width=1.5pt] (0,0) -- (1,0) -- (1,1) -- (2,1) -- (3,1) -- (3,2) -- (3,3);
\end{tikzpicture}
\begin{tikzpicture}[scale = 0.8]
  \draw[step=1cm,gray,very thin] (-0.1,-0.1) grid (3.1,3.1);
  \draw [gray,very thin](0,0) -- (3,3);
  \filldraw (0,0) circle (.1cm);
  \filldraw (3,3) circle (.1cm);
  \draw[line width=1.5pt] (0,0) -- (1,0) -- (2,0) -- (2,1) -- (3,1) -- (3,2) -- (3,3);
\end{tikzpicture}
\begin{tikzpicture}[scale = 0.8]
  \draw[step=1cm,gray,very thin] (-0.1,-0.1) grid (3.1,3.1);
  \draw [gray,very thin](0,0) -- (3,3);
  \filldraw (0,0) circle (.1cm);
  \filldraw (3,3) circle (.1cm);
  \draw[line width=1.5pt] (0,0) -- (1,0) -- (2,0) -- (2,1) -- (2,2) -- (3,2) -- (3,3);
\end{tikzpicture}
\begin{tikzpicture}[scale = 0.8]
  \draw[step=1cm,gray,very thin] (-0.1,-0.1) grid (3.1,3.1);
  \draw [gray,very thin](0,0) -- (3,3);
  \filldraw (0,0) circle (.1cm);
  \filldraw (3,3) circle (.1cm);
  \draw[line width=1.5pt] (0,0) -- (1,0) -- (1,1) -- (2,1) -- (2,2) -- (3,2) -- (3,3);
\end{tikzpicture}
    \caption{Five possible Dyck paths for $n=3$.}
    \label{fig_Dyck_paths}
\end{figure}
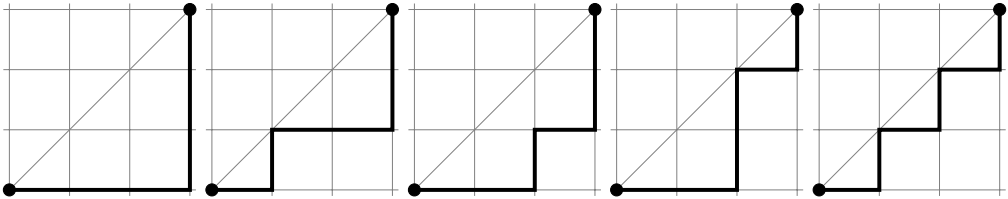

To better understand the number of figures in $\mathbf{M}^{(k,0)}$,
 we introduce the concept of Dyck paths.
In combinatorics, a Dyck path \cite{deutsch1999dyck} refers to a path on the plane satisfying the following properties:
\begin{itemize}
    \item the path starts from $(0,0)$ and ends at $(n,n)$;
    \item each step of the path is either $(1,0)$ or $(0,1)$, i.e., only upward or rightward movements are allowed;
    \item the entire path lies below (may touch) the line $y=x$.
\end{itemize}
For example, \cref{fig_Dyck_paths} shows 
 that there are in total 5 Dyck paths when $n=3$.
Comparing \cref{fig_Dyck_paths} and \cref{fig_M40}, one can see that
 the profiles of the bottom right boundaries in \cref{fig_M40}
 match exactly the Dyck paths given in \cref{fig_Dyck_paths}.
Hence, the number of terms in $\mathbf{M}^{(k,0)}$
 is identical to the number of Dyck paths when $n=k-1$.
The number of Dyck paths from $(0,0)$ to $(n,n)$,
 which is well-studied in combinatorics,
 is called Catalan numbers $C_n$ \cite{deutsch1999dyck,harris2008combinatorics}
 with the following explicit expression:
\begin{equation*}
    C_n = \frac{1}{n+1} \binom{2n}{n}
    = \frac{(2n)!}{(n+1)!n!}.
\end{equation*}
By Stirling's formula, the Catalan number has the following asymptotic form for large $n$:
\begin{equation}
\label{eq_asymptotic}
    C_n \sim \frac{4^n}{\sqrt{\pi}n^{3/2}}.
\end{equation}
To better understand the definition of $\mathbf{M}^{(k,0)}$,
 we take $\mathbf{M}^{(4,0)}$ as an example.
We can write down the explicit expression of $\M{4}{0}$ by \cref{fig_M40}:
\begin{equation}
\label{eq_M40}
\begin{split}
    \M{4}{0} =& \sum_{\sigma_1^\pm,\sigma_2^\pm,\sigma_3^\pm=\pm1}
    \A{4}{3} \A{3}{2} \A{2}{1} \A{1}{0} \\
    &\times\left[
    \F{2}{0}\F{3}{1}\F{3}{0}\F{4}{2}\F{4}{1}(\F{4}{0}-1) \right.\\
    &+ \F{3}{1}\F{4}{2}(\F{2}{0}-1)(\F{4}{2}-1) \\
    &+ \F{2}{0}\F{3}{1}\F{4}{2}(\F{3}{0}-1)(\F{4}{1}-1) \\
    &+ \F{2}{0}\F{3}{1}(\F{3}{0}-1)(\F{4}{2}-1) \\
    &+ \left.(\F{2}{0}-1)(\F{3}{1}-1)(\F{4}{2}-1)
    \right].
\end{split}
\end{equation}
If we directly follow the definition \cref{eq_M40}, the computational cost for the terms in the square brackets
 is the sum of the number of $F$ or $F-1$ in the product in each figure.
For the first figure (first term in the bracket),
 the computational cost is 
 counted as $6$ 
 and for the second term, the computational cost is $4$.
By considering all the terms, the computational cost for the matrix $\mathbf{M}^{(4,0)}$ is $2^{2\times 5}\times (6+4+5+4+3)$
 where $2^{2\times 5}$ comes from the summation over all possible paths.
The sum $(6+4+5+4+3)$ comes from the bracket in \cref{eq_M40}.
One can also find the number $(6+4+5+4+3)$ from the Dyck paths in \cref{fig_Dyck_paths}
 by counting the squares and triangles
 above each path and under the line $y=x$.
For example, in the first Dyck path, there are 3 squares and 3 triangles (3+3=6)
 above the path and below the line $y=x$.
In general,
for each term in $\mathbf{M}^{(k,0)}$,
 the computational cost is proportional 
 to the number of $F$ or $F-1$ in the product.
We can also use the total number of rectangles and triangles above 
 the corresponding Dyck path and below the line $y=x$
 to estimate the computational cost for the term.
Therefore, the computational cost for $\mathbf{M}^{(k,0)}$ for $k=1,\dots,\dk$ is estimated by
\begin{equation}
    \label{eq_complexity_SMatPI}
    \sum_{k=1}^\dk 2^{2(k+1)} S_k
\end{equation}
where $S_k$ is the total number of rectangles and triangles above all Dyck paths.
For example, $S_4 = 6+4+5+4+3 = 22$.
\begin{figure}
    \centering
    \begin{subfigure}{0.48\textwidth}
    \centering
        \includegraphics[width=\textwidth]{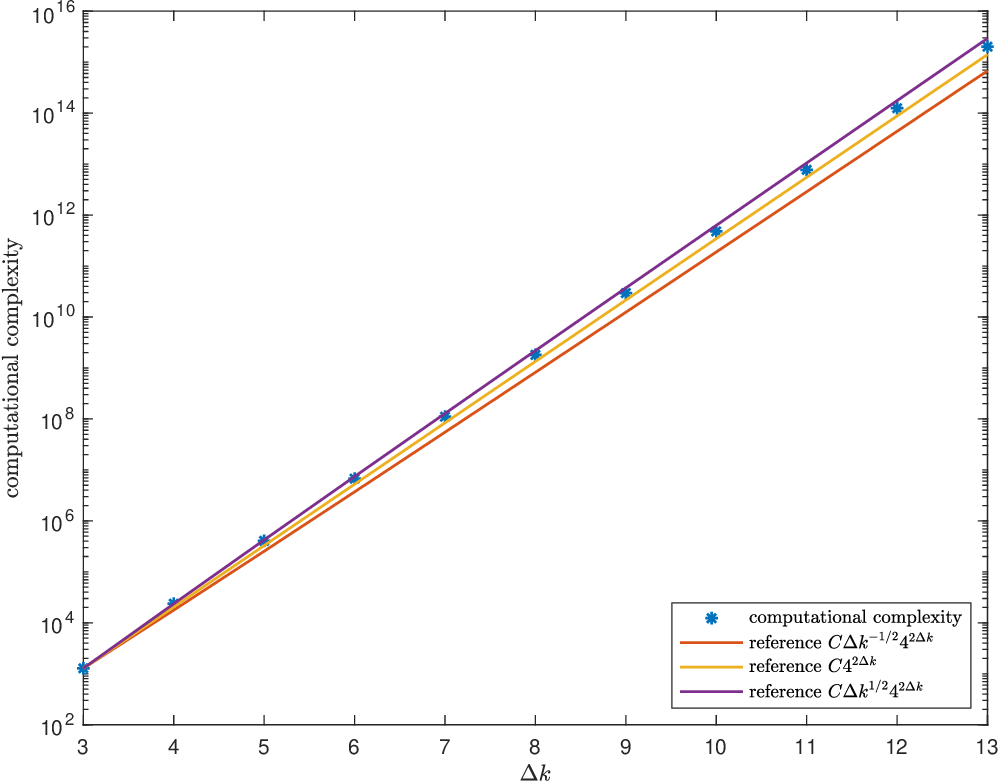}
        \caption{Computational complexity for SMatPI matrices.}
    \end{subfigure}%
    ~ 
    \begin{subfigure}{0.48\textwidth}
    \centering
        \includegraphics[width=\textwidth]{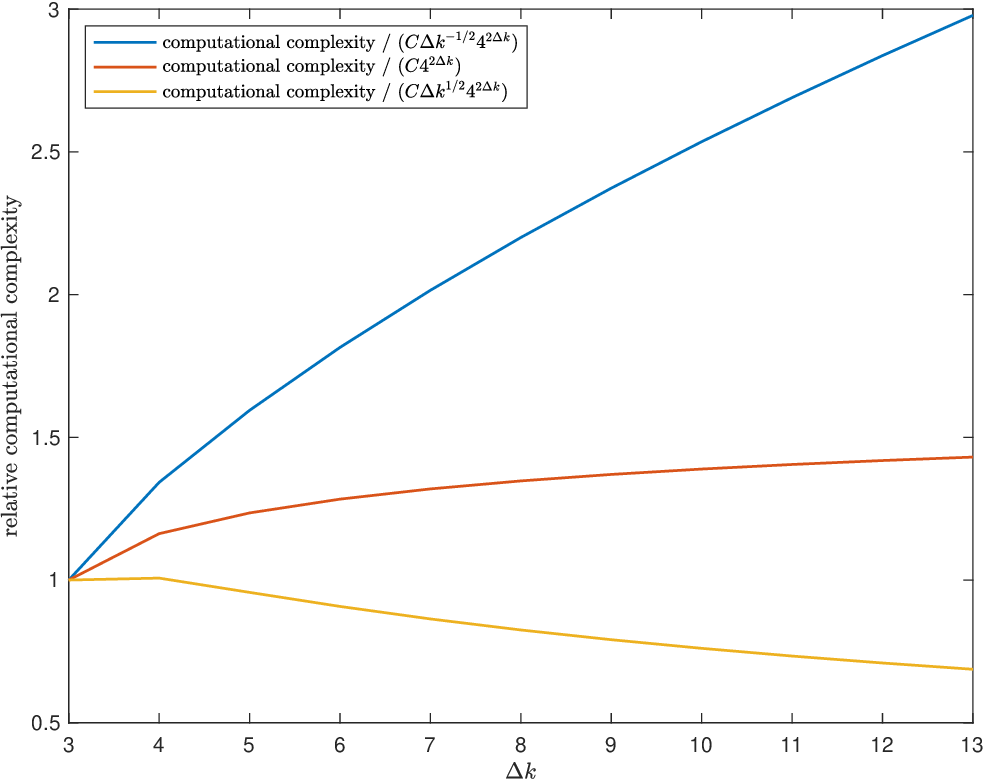}
        \caption{Relative computational complexity.}
    \end{subfigure}
    \caption{The growth of computational complexity for SMatPI matrices.}
    \label{fig_computational_cost_SMatPI}
\end{figure}
We now estimate the order of $S_k$
 by looking at the computational cost of each figure.
For the figures in $\mathbf{M}^{(k,0)}$ with a fixed path,
 the computational cost varies from $O(k)$ to $O(k^2)$.
For example, if the corresponding Dyck path has as many ``stairsteps'' as possible,
 the computational cost is $O(k)$ (see the last Dyck path in \cref{fig_Dyck_paths});
 if the Dyck path has only one ``stairstep'', the computational cost is $O(k^2)$ (see the first Dyck path in \cref{fig_Dyck_paths}).
As for the other Dyck paths,
 the corresponding computational cost is in between $O(k)$ and $O(k^2)$.
Therefore, we conclude that
\begin{equation*}
    k C_{k-1} \lesssim S_k \lesssim k^2 C_{k-1}
\end{equation*}
From the asymptotic expression of $C_n$ \cref{eq_asymptotic},
 we conclude that the computational cost \cref{eq_complexity_SMatPI}
 has upper bound $O(\dk^{1/2}4^{2\dk})$ and lower bound $O(\dk^{-1/2}4^{2\dk})$.
In \cref{fig_computational_cost_SMatPI}, 
 we plot three curves of $O(\dk^{\alpha} 4^{2\dk})$ with $\alpha = -1/2, 0, 1/2$, 
 which suggests that the asymptotic behavior of the actual computational cost \eqref{eq_complexity_SMatPI} is most likely to be $O(4^{2\dk})$.
Although it is not surprising that the computational cost of a path-integral method is exponential with respect to the memory length, 
the average computational cost for each path does not need to be also exponential even if we follow the definition of the SMatPI matrices. 
In the next two sections, we will introduce an approach that only requires polynomial time for each path.

\section{Recurrence relation for the kernel matrices}
\label{sec_fast_algorithm}
As analyzed in the previous section, 
 the main computational cost of the SMatPI 
 lies on the computation of matrices $\mathbf{M}^{(k,0)}$ or $\mathbf{M}^{(k+1,1)}$
 for $k=1,\dots,\Delta k$.
Inspired by the relation between the Catalan number and Catalan's triangle in combinatorics, 
 we design a recurrence algorithm for the computation of memory kernel
 $\mathbf{M}^{(k,0)}$ for $k=1,\dots,\Delta k$
 to accelerate the SMatPI. 
Some details about Catalan's triangle
 are available in the appendix.

For simplicity, we use the notation $\Mcal{k}{0}$ 
 ($\Mcal{k+1}{1}$) to 
 denote the term in $\M{k}{0}$ ($\M{k+1}{1}$) for a specific
 path $\boldsymbol{\sigma}^\pm = (\sigma_0^\pm, \dots, \sigma_{\dk}^{\pm})$, so that
\begin{equation}
\label{eq_M_split_to_Mcal}
    \M{k}{0} = 
    \sum_{\sigma_1^\pm,\dots,\sigma_{k-1}^\pm=\pm1}
    \Mcal{k}{0}.
\end{equation}
In particular,
\begin{equation}
\label{eq_initial_condition}
    \Mcal{1}{0} = 
    \M{1}{0} =  
    \G{1}{0}\F{1}{1}\F{1}{0}\F{0}{0}.
\end{equation}
In the notation $\Mcal{k}{0}$, the subscript $\sigma(0{:}k)$ indicates that this variable only relies on the zeroth component to the $k$th component of $\boldsymbol{\sigma}$, {i.e.}, $(\sigma_0^\pm,\dots,\sigma_k^\pm)$.
For simplicity, we only discuss $\Mcal{k}{0}$ here
 and the idea is exactly the same for the computation of $\Mcal{k+1}{1}$.
To reduce the computational cost, we will further split the variable $\Mcal{k}{0}$ into $k-1$ parts.
For example, the five terms in \cref{fig_M40} is further split into 3 categories based on the location of the shaded rectangle in the last column:
\begin{equation}
\label{eq_Mcal40}
    \Mcal{4}{0} = (\F{4}{0}-1)\Mscr{4}{0}{0}
    + (\F{4}{1}-1)\Mscr{4}{0}{1}
    + (\F{4}{2}-1)\Mscr{4}{0}{2},
\end{equation}
which can be diagrammatically written as
\begin{equation*}
\begin{split}
\Mcal{4}{0} =& (\F{4}{0}-1)\times
\left(
\begin{tikzpicture}[baseline={(current bounding box.center)},scale=0.4]
	\draw (0,0) -- (0.5,0.5);
	\draw (0.5,0.5) -- (0.5,0);
	\draw (0.5,0) -- (0,0);
	\draw (0.5,0) -- (0.5,0.5);
	\draw (0.5,0.5) -- (1.5,0.5);
	\draw (1.5,0.5) -- (1.5,0);
	\draw (1.5,0) -- (0.5,0);
	\draw (0.5,0.5) -- (1.5,1.5);
	\draw (1.5,1.5) -- (1.5,0.5);
	\draw (1.5,0.5) -- (0.5,0.5);
	\draw (1.5,0) -- (1.5,0.5);
	\draw (1.5,0.5) -- (2.5,0.5);
	\draw (2.5,0.5) -- (2.5,0);
	\draw (2.5,0) -- (1.5,0);
	\draw (1.5,0.5) -- (1.5,1.5);
	\draw (1.5,1.5) -- (2.5,1.5);
	\draw (2.5,1.5) -- (2.5,0.5);
	\draw (2.5,0.5) -- (1.5,0.5);
	\draw (1.5,1.5) -- (2.5,2.5);
	\draw (2.5,2.5) -- (2.5,1.5);
	\draw (2.5,1.5) -- (1.5,1.5);
	\draw (2.5,0) -- (2.5,0.5);
	\draw (2.5,0.5) -- (3.5,0.5);
	\draw (3.5,0.5) -- (3.5,0);
	\draw (3.5,0) -- (2.5,0);
	\draw (2.5,0.5) -- (2.5,1.5);
	\draw (2.5,1.5) -- (3.5,1.5);
	\draw (3.5,1.5) -- (3.5,0.5);
	\draw (3.5,0.5) -- (2.5,0.5);
	\draw (2.5,1.5) -- (2.5,2.5);
	\draw (2.5,2.5) -- (3.5,2.5);
	\draw (3.5,2.5) -- (3.5,1.5);
	\draw (3.5,1.5) -- (2.5,1.5);
	\draw (2.5,2.5) -- (3.5,3.5);
	\draw (3.5,3.5) -- (3.5,2.5);
	\draw (3.5,2.5) -- (2.5,2.5);
	\draw (3.5,0.5) -- (3.5,1.5);
	\draw (3.5,1.5) -- (4.5,1.5);
	\draw (4.5,1.5) -- (4.5,0.5);
	\draw (4.5,0.5) -- (3.5,0.5);
	\draw (3.5,1.5) -- (3.5,2.5);
	\draw (3.5,2.5) -- (4.5,2.5);
	\draw (4.5,2.5) -- (4.5,1.5);
	\draw (4.5,1.5) -- (3.5,1.5);
	\draw (3.5,2.5) -- (3.5,3.5);
	\draw (3.5,3.5) -- (4.5,3.5);
	\draw (4.5,3.5) -- (4.5,2.5);
	\draw (4.5,2.5) -- (3.5,2.5);
	\draw (3.5,3.5) -- (4.5,4.5);
	\draw (4.5,4.5) -- (4.5,3.5);
	\draw (4.5,3.5) -- (3.5,3.5);
\end{tikzpicture}
\right)
+ (\F{4}{1}-1)\times \left(
\begin{tikzpicture}[baseline={(current bounding box.center)},scale=0.4]
	\draw (0,0) -- (0.5,0.5);
	\draw (0.5,0.5) -- (0.5,0);
	\draw (0.5,0) -- (0,0);
	\draw (0.5,0) -- (0.5,0.5);
	\draw (0.5,0.5) -- (1.5,0.5);
	\draw (1.5,0.5) -- (1.5,0);
	\draw (1.5,0) -- (0.5,0);
	\draw (0.5,0.5) -- (1.5,1.5);
	\draw (1.5,1.5) -- (1.5,0.5);
	\draw (1.5,0.5) -- (0.5,0.5);
	\draw (1.5,0) -- (1.5,0.5);
	\draw (1.5,0.5) -- (2.5,0.5);
	\draw (2.5,0.5) -- (2.5,0);
	\draw (2.5,0) -- (1.5,0);
	\draw (1.5,0.5) -- (1.5,1.5);
	\draw (1.5,1.5) -- (2.5,1.5);
	\draw (2.5,1.5) -- (2.5,0.5);
	\draw (2.5,0.5) -- (1.5,0.5);
	\draw (1.5,1.5) -- (2.5,2.5);
	\draw (2.5,2.5) -- (2.5,1.5);
	\draw (2.5,1.5) -- (1.5,1.5);
	\draw (2.5,0) -- (2.5,0.5);
	\draw (2.5,0.5) -- (3.5,0.5);
	\draw (3.5,0.5) -- (3.5,0);
	\draw (3.5,0) -- (2.5,0);
	\draw (2.5,0) -- (2.5,0);
	\draw (2.8,0) -- (2.5,0.3);
	\draw (3.1,0) -- (2.6,0.5);
	\draw (3.4,0) -- (2.9,0.5);
	\draw (3.5,0.2) -- (3.2,0.5);
	\draw (3.5,0.5) -- (3.5,0.5);
	\draw (2.5,0.5) -- (2.5,1.5);
	\draw (2.5,1.5) -- (3.5,1.5);
	\draw (3.5,1.5) -- (3.5,0.5);
	\draw (3.5,0.5) -- (2.5,0.5);
	\draw (2.5,1.5) -- (2.5,2.5);
	\draw (2.5,2.5) -- (3.5,2.5);
	\draw (3.5,2.5) -- (3.5,1.5);
	\draw (3.5,1.5) -- (2.5,1.5);
	\draw (2.5,2.5) -- (3.5,3.5);
	\draw (3.5,3.5) -- (3.5,2.5);
	\draw (3.5,2.5) -- (2.5,2.5);
	\draw (3.5,1.5) -- (3.5,2.5);
	\draw (3.5,2.5) -- (4.5,2.5);
	\draw (4.5,2.5) -- (4.5,1.5);
	\draw (4.5,1.5) -- (3.5,1.5);
	\draw (3.5,2.5) -- (3.5,3.5);
	\draw (3.5,3.5) -- (4.5,3.5);
	\draw (4.5,3.5) -- (4.5,2.5);
	\draw (4.5,2.5) -- (3.5,2.5);
	\draw (3.5,3.5) -- (4.5,4.5);
	\draw (4.5,4.5) -- (4.5,3.5);
	\draw (4.5,3.5) -- (3.5,3.5);
\end{tikzpicture}
+
\begin{tikzpicture}[baseline={(current bounding box.center)},scale=0.4]
	\draw (0,0) -- (0.5,0.5);
	\draw (0.5,0.5) -- (0.5,0);
	\draw (0.5,0) -- (0,0);
	\draw (0.5,0) -- (0.5,0.5);
	\draw (0.5,0.5) -- (1.5,0.5);
	\draw (1.5,0.5) -- (1.5,0);
	\draw (1.5,0) -- (0.5,0);
	\draw (0.5,0.5) -- (1.5,1.5);
	\draw (1.5,1.5) -- (1.5,0.5);
	\draw (1.5,0.5) -- (0.5,0.5);
	\draw (1.5,0) -- (1.5,0.5);
	\draw (1.5,0.5) -- (2.5,0.5);
	\draw (2.5,0.5) -- (2.5,0);
	\draw (2.5,0) -- (1.5,0);
	\draw (1.5,0) -- (1.5,0);
	\draw (1.8,0) -- (1.5,0.3);
	\draw (2.1,0) -- (1.6,0.5);
	\draw (2.4,0) -- (1.9,0.5);
	\draw (2.5,0.2) -- (2.2,0.5);
	\draw (2.5,0.5) -- (2.5,0.5);
	\draw (1.5,0.5) -- (1.5,1.5);
	\draw (1.5,1.5) -- (2.5,1.5);
	\draw (2.5,1.5) -- (2.5,0.5);
	\draw (2.5,0.5) -- (1.5,0.5);
	\draw (1.5,1.5) -- (2.5,2.5);
	\draw (2.5,2.5) -- (2.5,1.5);
	\draw (2.5,1.5) -- (1.5,1.5);
	\draw (2.5,0.5) -- (2.5,1.5);
	\draw (2.5,1.5) -- (3.5,1.5);
	\draw (3.5,1.5) -- (3.5,0.5);
	\draw (3.5,0.5) -- (2.5,0.5);
	\draw (2.5,1.5) -- (2.5,2.5);
	\draw (2.5,2.5) -- (3.5,2.5);
	\draw (3.5,2.5) -- (3.5,1.5);
	\draw (3.5,1.5) -- (2.5,1.5);
	\draw (2.5,2.5) -- (3.5,3.5);
	\draw (3.5,3.5) -- (3.5,2.5);
	\draw (3.5,2.5) -- (2.5,2.5);
	\draw (3.5,1.5) -- (3.5,2.5);
	\draw (3.5,2.5) -- (4.5,2.5);
	\draw (4.5,2.5) -- (4.5,1.5);
	\draw (4.5,1.5) -- (3.5,1.5);
	\draw (3.5,2.5) -- (3.5,3.5);
	\draw (3.5,3.5) -- (4.5,3.5);
	\draw (4.5,3.5) -- (4.5,2.5);
	\draw (4.5,2.5) -- (3.5,2.5);
	\draw (3.5,3.5) -- (4.5,4.5);
	\draw (4.5,4.5) -- (4.5,3.5);
	\draw (4.5,3.5) -- (3.5,3.5);
\end{tikzpicture}
\right) \\
+& (\F{4}{2}-1)\times \left(
\begin{tikzpicture}[baseline={(current bounding box.center)},scale=0.4]
	\draw (0,0) -- (0.5,0.5);
	\draw (0.5,0.5) -- (0.5,0);
	\draw (0.5,0) -- (0,0);
	\draw (0.5,0) -- (0.5,0.5);
	\draw (0.5,0.5) -- (1.5,0.5);
	\draw (1.5,0.5) -- (1.5,0);
	\draw (1.5,0) -- (0.5,0);
	\draw (0.5,0.5) -- (1.5,1.5);
	\draw (1.5,1.5) -- (1.5,0.5);
	\draw (1.5,0.5) -- (0.5,0.5);
	\draw (1.5,0) -- (1.5,0.5);
	\draw (1.5,0.5) -- (2.5,0.5);
	\draw (2.5,0.5) -- (2.5,0);
	\draw (2.5,0) -- (1.5,0);
	\draw (1.5,0.5) -- (1.5,1.5);
	\draw (1.5,1.5) -- (2.5,1.5);
	\draw (2.5,1.5) -- (2.5,0.5);
	\draw (2.5,0.5) -- (1.5,0.5);
	\draw (1.5,1.5) -- (2.5,2.5);
	\draw (2.5,2.5) -- (2.5,1.5);
	\draw (2.5,1.5) -- (1.5,1.5);
	\draw (2.5,0) -- (2.5,0.5);
	\draw (2.5,0.5) -- (3.5,0.5);
	\draw (3.5,0.5) -- (3.5,0);
	\draw (3.5,0) -- (2.5,0);
	\draw (2.5,0) -- (2.5,0);
	\draw (2.8,0) -- (2.5,0.3);
	\draw (3.1,0) -- (2.6,0.5);
	\draw (3.4,0) -- (2.9,0.5);
	\draw (3.5,0.2) -- (3.2,0.5);
	\draw (3.5,0.5) -- (3.5,0.5);
	\draw (2.5,0.5) -- (2.5,1.5);
	\draw (2.5,1.5) -- (3.5,1.5);
	\draw (3.5,1.5) -- (3.5,0.5);
	\draw (3.5,0.5) -- (2.5,0.5);
	\draw (2.5,1.5) -- (2.5,2.5);
	\draw (2.5,2.5) -- (3.5,2.5);
	\draw (3.5,2.5) -- (3.5,1.5);
	\draw (3.5,1.5) -- (2.5,1.5);
	\draw (2.5,2.5) -- (3.5,3.5);
	\draw (3.5,3.5) -- (3.5,2.5);
	\draw (3.5,2.5) -- (2.5,2.5);
	\draw (3.5,2.5) -- (3.5,3.5);
	\draw (3.5,3.5) -- (4.5,3.5);
	\draw (4.5,3.5) -- (4.5,2.5);
	\draw (4.5,2.5) -- (3.5,2.5);
	\draw (3.5,3.5) -- (4.5,4.5);
	\draw (4.5,4.5) -- (4.5,3.5);
	\draw (4.5,3.5) -- (3.5,3.5);
\end{tikzpicture}
+
\begin{tikzpicture}[baseline={(current bounding box.center)},scale=0.4]
	\draw (0,0) -- (0.5,0.5);
	\draw (0.5,0.5) -- (0.5,0);
	\draw (0.5,0) -- (0,0);
	\draw (0.5,0) -- (0.5,0.5);
	\draw (0.5,0.5) -- (1.5,0.5);
	\draw (1.5,0.5) -- (1.5,0);
	\draw (1.5,0) -- (0.5,0);
	\draw (0.5,0.5) -- (1.5,1.5);
	\draw (1.5,1.5) -- (1.5,0.5);
	\draw (1.5,0.5) -- (0.5,0.5);
	\draw (1.5,0) -- (1.5,0.5);
	\draw (1.5,0.5) -- (2.5,0.5);
	\draw (2.5,0.5) -- (2.5,0);
	\draw (2.5,0) -- (1.5,0);
	\draw (1.5,0) -- (1.5,0);
	\draw (1.8,0) -- (1.5,0.3);
	\draw (2.1,0) -- (1.6,0.5);
	\draw (2.4,0) -- (1.9,0.5);
	\draw (2.5,0.2) -- (2.2,0.5);
	\draw (2.5,0.5) -- (2.5,0.5);
	\draw (1.5,0.5) -- (1.5,1.5);
	\draw (1.5,1.5) -- (2.5,1.5);
	\draw (2.5,1.5) -- (2.5,0.5);
	\draw (2.5,0.5) -- (1.5,0.5);
	\draw (1.5,1.5) -- (2.5,2.5);
	\draw (2.5,2.5) -- (2.5,1.5);
	\draw (2.5,1.5) -- (1.5,1.5);
	\draw (2.5,0.5) -- (2.5,1.5);
	\draw (2.5,1.5) -- (3.5,1.5);
	\draw (3.5,1.5) -- (3.5,0.5);
	\draw (3.5,0.5) -- (2.5,0.5);
	\draw (2.5,0.5) -- (2.5,0.5);
	\draw (2.9,0.5) -- (2.5,0.9);
	\draw (3.3,0.5) -- (2.5,1.3);
	\draw (3.5,0.7) -- (2.7,1.5);
	\draw (3.5,1.1) -- (3.1,1.5);
	\draw (3.5,1.5) -- (3.5,1.5);
	\draw (2.5,1.5) -- (2.5,2.5);
	\draw (2.5,2.5) -- (3.5,2.5);
	\draw (3.5,2.5) -- (3.5,1.5);
	\draw (3.5,1.5) -- (2.5,1.5);
	\draw (2.5,2.5) -- (3.5,3.5);
	\draw (3.5,3.5) -- (3.5,2.5);
	\draw (3.5,2.5) -- (2.5,2.5);
	\draw (3.5,2.5) -- (3.5,3.5);
	\draw (3.5,3.5) -- (4.5,3.5);
	\draw (4.5,3.5) -- (4.5,2.5);
	\draw (4.5,2.5) -- (3.5,2.5);
	\draw (3.5,3.5) -- (4.5,4.5);
	\draw (4.5,4.5) -- (4.5,3.5);
	\draw (4.5,3.5) -- (3.5,3.5);
\end{tikzpicture}
\right).
\end{split}
\end{equation*}%
Note that the variables $\Mscr{4}{0}{j},j=0,1,2$ remove the shaded rectangles in the last column of the triangle, so that when we compute $\Mcal{4}{0}$,
 we need to assign each $\Mscr{4}{0}{j}$ a factor $(\F{4}{j}-1)$.
The reason is that
 without these shaded rectangles,
 we are free to add a shaded rectangle by multiplying $(\F{4}{j}-1)$ or add a normal rectangle by multiplying $\F{4}{j}$ when expanding the diagrams,
 which will be used later
 for the recurrence relation.

For general $k\geqslant 2$,
 $\Mcal{k}{0}$ is further split into $k-1$ parts,
 which are denoted by $\Mscr{k}{0}{j}$ for $j=0,\dots,k-1$.
Each $\Mscr{k}{0}{j}$ represents all the figures which originally have a shaded rectangle at $(k,j)$
although the shaded rectangle for each figure at the last column in
 $\Mscr{k}{0}{j}$ is then removed
 for convenience of the following discussion.
With the splitting,
 we can write $\Mcal{k}{0}$ as a linear combination of $\Mscr{k}{0}{j}$:
\begin{equation}
    \label{eq_Mk0_split}
    \Mcal{k}{0} = \sum_{j=0}^{k-2} (\F{k}{j}-1)\Mscr{k}{0}{j}.
\end{equation}
In \cref{eq_Mk0_split},
 the coefficients $(\F{k}{j}-1)$
 mean adding a shaded rectangle at the bottom of the last column.
Combining \cref{eq_M_split_to_Mcal,eq_Mk0_split},
 we have
\begin{equation}
\label{eq_M_split_to_Mscr}
    \M{k}{0} = \sum_{\sigma_1^\pm,\dots,\sigma_{k-1}^\pm=\pm1}
    \sum_{j=0}^{k-2} (\F{k}{j}-1) \Mscr{k}{0}{j}
\end{equation}
for $k\geqslant 2$.

The advantage of the splitting is that we can construct $\Mscr{k+1}{0}{\cdot}$
 using $\Mscr{k}{0}{\cdot}$, allowing us to
 compute $\Mscr{k}{0}{\cdot}$ recursively
 by adding the last column accordingly.
For example, $\Mscr{4}{0}{\cdot}$ can be computed from $\Mscr{3}{0}{\cdot}$ by adding the last column in the diagrams.
\begin{equation}
\label{eq_Mscr40}
    \begin{split}
        \Mscr{4}{0}{0} &= \F{3}{0}\Mscr{3}{0}{0} 
        \A{4}{3} \F{4}{2} \F{4}{1}; \\
        \Mscr{4}{0}{1} &= \left((\F{3}{0} - 1) \Mscr{3}{0}{0}
        + \F{3}{1} \Mscr{3}{0}{1}
        \right) \A{4}{3} \F{4}{2} ;\\
        \Mscr{4}{0}{2} &= \left(
        (\F{3}{0}-1) \Mscr{3}{0}{0} + (\F{3}{1}-1) \Mscr{3}{0}{1}
        \right)\A{4}{3}.
    \end{split}
\end{equation}
We may also represent \cref{eq_Mscr40} by the following
 diagrammatic equations.
\input{Mscr40.tex}
In general, the recurrence relation is given by
\begin{equation}
\label{eq_Mscr_recurrence}
    \Mscr{k+1}{0}{n} = 
    \left(\sum_{j=0}^{n-1}(\F{k}{j}-1)\Mscr{k}{0}{j}
    + \F{k}{n} \Mscr{k}{0}{n}\right)
    \A{k+1}{k} 
    \left(\prod_{l=n+1}^{k-1} \F{k+1}{l}\right)
\end{equation}
for $k = 2,\dots,\dk-1$ and $n = 0,\dots,k-1$.
Note that when $n=k-1$,
 the term $\Mscr{k}{0}{k-1}$ on the right hand side is regarded as 0.
In addition, we also have the following relation
 between $\Mscr{2}{0}{0}$ and $\Mcal{1}{0}$.
\begin{equation}
\label{eq_Mscr20_Mcal10}
    \Mscr{2}{0}{0} = \Mcal{1}{0} \A{2}{1}.
\end{equation}
Similarly, the matrices $\mathbf{M}^{(k+1,1)}$
for $k=1,\dots,\dk$ has the following recurrence relations:
\begin{align}
    &\Mcal{2}{1} = \G{2}{1} \F{2}{2}\F{2}{1}, 
    \label{eq_initial_21}\\
    &\Mscr{3}{1}{1} = \Mcal{2}{1} \A{3}{2}, 
    \label{eq_M21_M31}\\
    &\Mscr{k+1}{1}{n}
    = \left(
    \sum_{j=1}^{n-1} (\F{k}{j}-1) \Mscr{k}{1}{j}
    + \F{k}{n} \Mscr{k}{1}{n}
    \right)
    \A{k+1}{k}
    \left(
    \prod_{l=n+1}^{k-1} \F{k+1}{l}
    \right),
    \label{eq_Mscr_recurrence_1}
    \\
    &\M{k+1}{1} = \sum_{\sigma_2^\pm,\dots,\sigma_k^\pm = \pm 1}
    \sum_{j=1}^{k-1} 
    (\F{k+1}{j} - 1) \Mscr{1}{k+1}{j},
    \label{eq_split_M1_Mscr}
\end{align}
where $n=1,\dots,k-1$ for \cref{eq_Mscr_recurrence_1} and $\Mscr{k}{1}{k-1}=0$ on the right-hand side if $n=k-1$.
The algorithm based on the recurrence relations will be detailed in the next section.

\begin{remark}
In this recurrence relation \cref{eq_Mscr_recurrence},
 the number of diagrams in $\Mscr{k}{0}{j}$
 is exactly $T(k-2,j)$ in the Catalan's triangle (see appendix for some more details).
\Cref{eq_Mk0_split} coincides with the property \cref{eq_sumTnj}.
The idea of computing $\Mscr{k+1}{0}{\cdot}$ from $\Mscr{k}{0}{\cdot}$ in
 \cref{eq_Mscr_recurrence} comes from \cref{eq_recurrence}.
\end{remark}

\section{Numerical implementation by the t-SMatPI method}
\begin{figure}
\scriptsize \begin{tikzpicture}
    \filldraw [yellow,rounded corners] (0,0) rectangle +(-2.8,-0.6);
    \node at (-1.4,-0.3) {$\mathscr{M}_{\{----\},\{----\}}^{(3,0),[\cdot]}$};
    \node at (0.2,-0.3) {$\natural$};
    \filldraw [yellow,rounded corners] (0,-0.8) rectangle +(-2.8,-0.6);
    \node at (-1.4,-1.1) {$\mathscr{M}_{\{----\},\{---+\}}^{(3,0),[\cdot]}$};
    \node at (0.2,-1.1) {$\sharp$};
    \filldraw [pink,rounded corners] (0,-1.6) rectangle +(-2.8,-0.6);
    \node at (-1.4,-1.9) {$\mathscr{M}_{\{---+\},\{----\}}^{(3,0),[\cdot]}$};
    \node at (0.2,-1.9) {$\flat$};
    \filldraw [yellow,rounded corners] (0,-2.4) rectangle +(-2.8,-0.6);
    \node at (-1.4,-2.7) {$\mathscr{M}_{\{---+\},\{---+\}}^{(3,0),[\cdot]}$};
    \node at (0.2,-2.7) {$\heartsuit$};
    \filldraw [yellow,rounded corners] (0,-3.2) rectangle +(-2.8,-0.6);
    \node at (-1.4,-3.5) {$\mathscr{M}_{\{----\},\{--+-\}}^{(3,0),[\cdot]}$};
    \node at (0.2,-3.5) {$\clubsuit$};
    \filldraw [pink,rounded corners] (0,-4.0) rectangle +(-2.8,-0.6);
    \node at (-1.4,-4.3) {$\mathscr{M}_{\{----\},\{--++\}}^{(3,0),[\cdot]}$};
    \filldraw [pink,rounded corners] (0,-4.8) rectangle +(-2.8,-0.6);
    \node at (-1.4,-5.1) {$\mathscr{M}_{\{---+\},\{--+-\}}^{(3,0),[\cdot]}$};
    \filldraw [pink,rounded corners] (0,-5.6) rectangle +(-2.8,-0.6);
    \node at (-1.4,-5.9) {$\mathscr{M}_{\{---+\},\{--++\}}^{(3,0),[\cdot]}$};
    \filldraw [pink,rounded corners] (0,-6.4) rectangle +(-2.8,-0.6);
    \node at (-1.4,-6.7) {$\cdots$};
    \filldraw [pink,rounded corners] (0,-7.2) rectangle +(-2.8,-0.6);
    \node at (-1.4,-7.5) {$\mathscr{M}_{\{--+-\},\{--+-\}}^{(3,0),[\cdot]}$};
    \filldraw [pink,rounded corners] (0,-8.0) rectangle +(-2.8,-0.6);
    \node at (-1.4,-8.3) {$\mathscr{M}_{\{--+-\},\{--++\}}^{(3,0),[\cdot]}$};
    \filldraw [pink,rounded corners] (0,-8.8) rectangle +(-2.8,-0.6);
    \node at (-1.4,-9.1) {$\mathscr{M}_{\{--++\},\{--+-\}}^{(3,0),[\cdot]}$};
    \filldraw [yellow,rounded corners] (0,-9.6) rectangle +(-2.8,-0.6);
    \node at (-1.4,-9.9) {$\mathscr{M}_{\{--++\},\{--++\}}^{(3,0),[\cdot]}$};
    \node at (0.2,-9.9) {$\diamondsuit$};
    \filldraw [yellow,rounded corners] (0,-10.4) rectangle +(-2.8,-0.6);
    \node at (-1.4,-10.7) {$\mathscr{M}_{\{----\},\{-+--\}}^{(3,0),[\cdot]}$};
    \node at (0.2,-10.7) {$\spadesuit$};
    \filldraw [pink,rounded corners] (0,-11.2) rectangle +(-2.8,-0.6);
    \node at (-1.4,-11.5) {$\mathscr{M}_{\{----\},\{-+-+\}}^{(3,0),[\cdot]}$};
    \filldraw [pink,rounded corners] (0,-12.0) rectangle +(-2.8,-0.6);
    \node at (-1.4,-12.3) {$\mathscr{M}_{\{---+\},\{-+--\}}^{(3,0),[\cdot]}$};
    \filldraw [pink,rounded corners] (0,-12.8) rectangle +(-2.8,-0.6);
    \node at (-1.4,-13.1) {$\mathscr{M}_{\{---+\},\{-+-+\}}^{(3,0),[\cdot]}$};
    \filldraw [pink,rounded corners] (0,-13.6) rectangle +(-2.8,-0.6);
    \node at (-1.4,-13.9) {$\cdots$};
    \filldraw [pink,rounded corners] (0,-14.4) rectangle +(-2.8,-0.6);
    \node at (-1.4,-14.7) {$\cdots$};
    \filldraw [pink,rounded corners] (0,-15.2) rectangle +(-2.8,-0.6);
    \node at (-1.4,-15.5) {$\cdots$};
    \filldraw [pink,rounded corners] (0,-16.0) rectangle +(-2.8,-0.6);
    \node at (-1.4,-16.3) {$\cdots$};
    \filldraw [pink,rounded corners] (0,-16.8) rectangle +(-2.8,-0.6);
    \node at (-1.4,-17.1) {$\cdots$};
    \draw (-2.8,-0.3) -- (-3.8,-1.35);
    \draw (-2.8,-1.1) -- (-3.8,-1.45);
    \draw (-2.8,-1.9) -- (-3.8,-1.55);
    \draw (-2.8,-2.7) -- (-3.8,-1.65);
    \draw (-2.8,-3.5) -- (-3.8,-4.55);
    \draw (-2.8,-4.3) -- (-3.8,-4.65);
    \draw (-2.8,-5.1) -- (-3.8,-4.75);
    \draw (-2.8,-5.9) -- (-3.8,-4.85);
    \draw (-2.8,-6.7) -- (-3.8,-6.7);
    \draw (-2.8,-7.5) -- (-3.8,-8.55);
    \draw (-2.8,-8.3) -- (-3.8,-8.65);
    \draw (-2.8,-9.1) -- (-3.8,-8.75);
    \draw (-2.8,-9.9) -- (-3.8,-8.85);
    \draw (-2.8,-10.7) -- (-3.8,-11.75);
    \draw (-2.8,-11.5) -- (-3.8,-11.85);
    \draw (-2.8,-12.3) -- (-3.8,-11.95);
    \draw (-2.8,-13.1) -- (-3.8,-12.05);
    \draw (-2.8,-13.9) -- (-3.8,-13.9);
    \draw (-2.8,-14.7) -- (-3.8,-14.7);
    \draw (-2.8,-15.5) -- (-3.8,-15.5);
    \draw (-2.8,-16.3) -- (-3.8,-16.3);
    \draw (-2.8,-17.1) -- (-3.8,-17.1);
    \filldraw [pink,rounded corners] (-3.8,-1.2) rectangle +(-2.2,-0.6);
    \node at (-4.9,-1.5) {$\mathscr{M}_{\{---\},\{---\}}^{(2,0),[\cdot]}$};
    \filldraw [pink,rounded corners] (-3.8,-4.4) rectangle +(-2.2,-0.6);
    \node at (-4.9,-4.7) {$\mathscr{M}_{\{---\},\{--+\}}^{(2,0),[\cdot]}$};
    \filldraw [pink,rounded corners] (-3.8,-6.4) rectangle +(-2.2,-0.6);
    \node at (-4.9,-6.7) {$\mathscr{M}_{\{--+\},\{---\}}^{(2,0),[\cdot]}$};
    \filldraw [pink,rounded corners] (-3.8,-8.4) rectangle +(-2.2,-0.6);
    \node at (-4.9,-8.7) {$\mathscr{M}_{\{--+\},\{--+\}}^{(2,0),[\cdot]}$};
    \filldraw [pink,rounded corners] (-3.8,-11.6) rectangle +(-2.2,-0.6);
    \node at (-4.9,-11.9) {$\mathscr{M}_{\{---\},\{-+-\}}^{(2,0),[\cdot]}$};
    \filldraw [pink,rounded corners] (-3.8,-13.6) rectangle +(-2.2,-0.6);
    \node at (-4.9,-13.9) {$\mathscr{M}_{\{---\},\{-++\}}^{(2,0),[\cdot]}$};
    \filldraw [pink,rounded corners] (-3.8,-14.4) rectangle +(-2.2,-0.6);
    \node at (-4.9,-14.7) {$\mathscr{M}_{\{--+\},\{-+-\}}^{(2,0),[\cdot]}$};
    \filldraw [pink,rounded corners] (-3.8,-15.2) rectangle +(-2.2,-0.6);
    \node at (-4.9,-15.5) {$\mathscr{M}_{\{--+\},\{-++\}}^{(2,0),[\cdot]}$};
    \filldraw [pink,rounded corners] (-3.8,-16.0) rectangle +(-2.2,-0.6);
    \node at (-4.9,-16.3) {$\cdots$};
    \filldraw [pink,rounded corners] (-3.8,-16.8) rectangle +(-2.2,-0.6);
    \node at (-4.9,-17.1) {$\cdots$};
    \draw (-6.0,-1.5) -- (-7.0,-4.95);
    \draw (-6.0,-4.7) -- (-7.0,-5.05);
    \draw (-6.0,-6.7) -- (-7.0,-5.15);
    \draw (-6.0,-8.7) -- (-7.0,-5.25);
    \draw (-6.0,-11.9) -- (-7.0,-13.85);
    \draw (-6.0,-13.9) -- (-7.0,-13.95);
    \draw (-6.0,-14.7) -- (-7.0,-14.05);
    \draw (-6.0,-15.5) -- (-7.0,-14.15);
    \draw (-6.0,-16.3) -- (-7.0,-16.3);
    \draw (-6.0,-17.1) -- (-7.0,-17.1);
    \filldraw [pink,rounded corners] (-7.0,-4.8) rectangle +(-1.8,-0.6);
    \node at (-7.9,-5.1) {$\mathcal{M}_{\{--\},\{--\}}^{(1,0)}$};
    \filldraw [pink,rounded corners] (-7.0,-13.7) rectangle +(-1.8,-0.6);
    \node at (-7.9,-14.0) {$\mathcal{M}_{\{--\},\{-+\}}^{(1,0)}$};
    \filldraw [pink,rounded corners] (-7.0,-16.0) rectangle +(-1.8,-0.6);
    \node at (-7.9,-16.3) {$\mathcal{M}_{\{-+\},\{--\}}^{(1,0)}$};
    \filldraw [pink,rounded corners] (-7.0,-16.8) rectangle +(-1.8,-0.6);
    \node at (-7.9,-17.1) {$\mathcal{M}_{\{-+\},\{-+\}}^{(1,0)}$};
    \draw (-8.8,-5.1) -- (-10.5,-11.1);
    \draw (-8.8,-14.0) -- (-10.5,-11.1);
    \draw (-8.8,-16.3) -- (-10.5,-11.1);
    \draw (-8.8,-17.1) -- (-10.5,-11.1);
    \filldraw [pink] (-10.5,-11.1) circle (0.8);
    \node at (-10.5,-11.1) {$\{-\},\{-\}$};
\end{tikzpicture}
\caption{Illustration of the tree structure for the algorithm where $\dk=3$. 
In this figure, we only show the tree with $\sigma_0 = (-1,-1)$.
There are three other trees with $\sigma_0=(-1,+1),\sigma_0=(+1,-1),\sigma_0=(+1,+1)$,
 which are omitted in this figure.}
\label{fig_tree}
\end{figure}
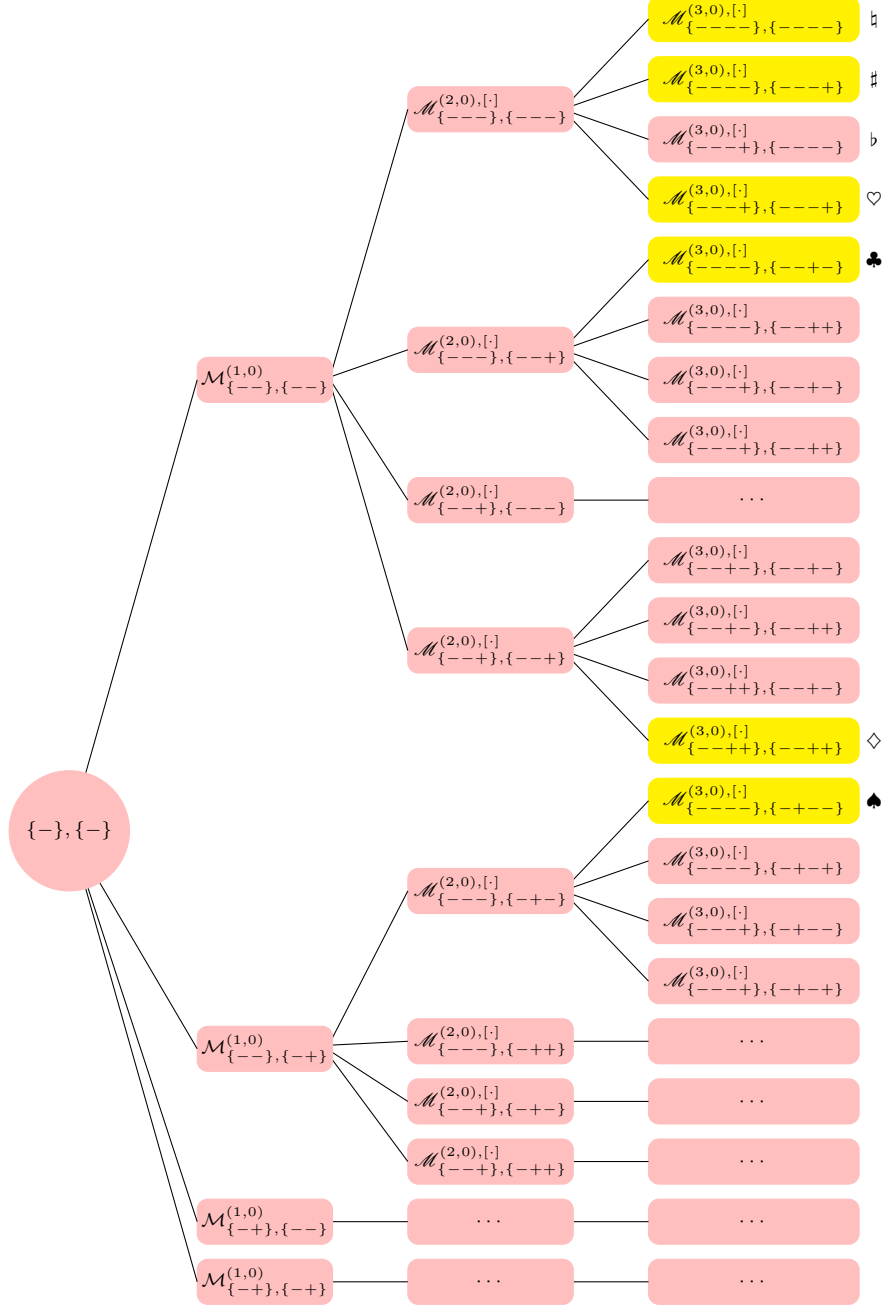
\label{sec_implementation}
The recurrence relation \cref{eq_Mscr_recurrence} applies for every path $\boldsymbol{\sigma}$.
If the memory length is $\dk$, there are $2^{2\dk+2}$ different paths.
Below, we will introduce a method that traverses a tree while computing the values of $\M{k}{0}$ so that the implementation
 does not require exponentially growing memory.
Since the method is the implementation of the small matrix decomposition path integral based on a tree structure, we call
 our method ``t-SMatPI'' where ``t'' stands for the tree structure.

The main purpose of t-SMatPI is to accelerate the computation
 of $\M{k}{0}$ for $k=1,\dots,\dk$.
According to \cref{eq_M_split_to_Mscr}, this requires us to compute the path-dependent quantities $\Mscr{k}{0}{n}$.
Note that the path $\boldsymbol{\sigma} = (\sigma_0,\sigma_1,\dots,\sigma_{\dk})$
 can be regarded as a time sequence 
 where each $\sigma_j = (\sigma_j^+,\sigma_j^-)$
 has 4 possible choices in the spin-boson model.
Therefore, all possible discrete paths can form a quadtree structure as shown in \cref{fig_tree}.
The tree in \cref{fig_tree} begins with a root node including only the first component of the path $\sigma_0$. Such a path has four possibilities: $\sigma_0 = (-1,-1), (-1,+1), (+1,-1), (+1,+1)$. There are thus four trees we need to consider in the algorithm, and we plot only one of them as an example. The root node has four children, including four possible extensions of the path $\sigma_0$ to $(\sigma_0, \sigma_1)$.
Similarly, each node on this level has four children including the four possible extensions of the path $(\sigma_0, \sigma_1)$ to $(\sigma_0, \sigma_1, \sigma_2)$, and such a structure continues until the length of the paths reaches $\dk$.
From the first level (the child nodes of the root node), each node includes one value $\Mcal{1}{0}$.
For other levels, each node includes $k-1$ values $\Mscr{k}{0}{n}$ for $n=0,\dots,k-2$, where $k$ corresponds to the depth of the node, and $\sigma(0{:}k)$ corresponds to a section of the path.
Based on \cref{eq_M_split_to_Mscr},
 the value of $\M{k}{0}$
 is actually the weighted sum of all the values
 in the $k$th level in \cref{fig_tree}.
 
In the quadtree, each node can be represented by a section of the path $\boldsymbol{\sigma}$, and the quantities $\mathcal{M}_{\sigma(0:1)}^{(1,0)}$ or
 $\Mscr{k}{0}{\cdot}$ that we place in each node depend only on this particular section of the path, i.e., $(\sigma_0,\dots,\sigma_k)$.
Moreover, the recurrence relation of $\mathscr{M}$ in \cref{eq_Mscr_recurrence} also fits the tree structure seamlessly.
It can be observed that in \cref{eq_Mscr_recurrence}, the quantity on the left-hand side ($\Mscr{k+1}{0}{\cdot}$) is always stored in the child of the node containing the quantities on the right-hand side ($\Mscr{k}{0}{\cdot}$),
 meaning that the values of the quantities in every node can be computed from its parent node.
Thus, by a depth-first traversal of the tree, all the quantities can be computed in order.
Note that although we draw a tree for illustration,
 there is no need to maintain such a data structure in the implementation.
The reason is that we only care about the weighted sum of the values stored in the tree,
 so that as long as the values stored in a node have been used to generate all its children and are added to the memory kernel matrix according to \cref{eq_M_split_to_Mscr},
 all the information in the node can be disposed to prevent the memory cost from increasing exponentially with respect to $\dk$.
In what follows, we will give an example to show how the memory-efficient algorithm is designed.

Consider the case $\dk=3$, for which one of the quadtrees is plotted in \cref{fig_tree}.
The purpose of the algorithm is to compute the matrices $\mathbf{M}^{(1,0)}$, $\mathbf{M}^{(2,0)}$ and $\mathbf{M}^{(3,0)}$ via a traversal of the tree.
As the elements of these matrices are represented by summations (see \cref{eq_M_split_to_Mscr}), we
 set them as zero in the beginning of the algorithm,
 after which we can start the depth-first tree traversal.
Reaching the first leaf node (labeled with $\natural$) is straightforward.
We just need to start from the node $\mathcal{M}_{\{--\},\{--\}}^{(1,0)}$  by the ``initial condition'' \cref{eq_initial_condition},
 and then compute $\mathscr{M}_{\{---\},\{---\}}^{(2,0),[\cdot]}$ according to \cref{eq_Mscr20_Mcal10}, which then leads to the leaf node $\natural$ according to \cref{eq_Mscr_recurrence}.
In our implementation, we also maintain a ``path variable'' $\boldsymbol{\sigma}$ to denote the subscripts in the nodes.
From the root node to the leaf node $\natural$, the value of $\boldsymbol{\sigma}$ grows from ``$\{-\},\{-\}$'' to ``$\{{-}{-}{-}{-}\},\{{-}{-}{-}{-}\}$''.
After computing all the quantities in these three nodes,
 we add them to their corresponding matrices $\mathbf{M}^{(k,0)}$ based on \cref{eq_M_split_to_Mcal} for $k=1$ 
 or \cref{eq_M_split_to_Mscr} for $k=2$ and $3$.
Afterwards, the information in the leaf node $\natural$ is no longer useful and can be discarded, but the information in its parent and grandparent must still stay in the memory since they have been used only to generate one of their children.
Then we can proceed with the traversal and
 move to the next node (labeled with $\sharp$).
Compared with the path for the previous node $\natural$,
 switching to the new node $\sharp$ only requires changing the last state of the path $\sigma_3$.
Similarly, the information in the node $\sharp$ can also be computed from its parent according to \cref{eq_M_split_to_Mscr}, and afterwards we can add them to $\mathbf{M}^{(3,0)}$ and then discard their values. Again, the information in its parent and grandparent nodes must stay in the memory.
In the implementation, when switching from $\natural$ to $\sharp$,
 we can reuse the memory allocated to the node $\natural$ and simply replace the values 
 by the quantities in the node $\sharp$ 
 so that the memory cost will not increase.
The same trick can be applied to the next two nodes $\flat$ and $\heartsuit$. 
By the rules of the depth-first traversal, the node to be visited after $\heartsuit$ should be the one including $\mathscr{M}_{\{---\},\{--+\}}^{(2,0),[\cdot]}$, whose values can be computed by  \cref{eq_Mscr20_Mcal10} since we still have its parent $\mathcal{M}_{\{--\},\{--\}}^{(1,0)}$ stored in the memory.
Note that once $\heartsuit$ is visited, the values of $\mathscr{M}_{\{---\},\{---\}}^{(2,0),[\cdot]}$, stored in the sibling of the current node, is no longer useful.
This allows us to write the values $\mathscr{M}_{\{---\},\{--+\}}^{(2,0),[\cdot]}$ into the memory allocated for $\mathscr{M}_{\{---\},\{---\}}^{(2,0),[\cdot]}$, so that the total memory cost, again, does not increase.
We would also like to remind the readers that the values of $\mathscr{M}_{\{---\},\{--+\}}^{(2,0),[\cdot]}$ should be added to $\mathbf{M}^{(2,0)}$ according to  \cref{eq_M_split_to_Mscr}.
Subsequently, the quantities in the node $\clubsuit$ can be evaluated according to the recurrence relation \cref{eq_Mscr_recurrence} and added to $\mathbf{M}^{(3,0)}$ according to \cref{eq_M_split_to_Mscr}, and this still requires no extra memory since the memory once used to store $\heartsuit$ is now available.

In general, during the traversal of the tree, we never need to allocate new memory once the first leaf node is visited.
For each layer of the tree, all the nodes can share the same memory location to store their information.
For example, after computing $\mathscr{M}_{\{--++\},\{--+-\}}^{(3,0),[\cdot]}$ (the node $\diamondsuit$),
 the subsequent nodes are $\mathcal{M}_{\{--\},\{-+\}}^{(1,0)}$, $\mathscr{M}_{\{---\},\{-+-\}}^{(2,0),[\cdot]}$ and $\mathscr{M}_{\{----\},\{-++-\}}^{(3,0),[\cdot]}$ (the path from the root to the node $\spadesuit$).
These values will take up the memory used for $\mathcal{M}_{\{--\},\{--\}}^{(1,0)}$, $\mathscr{M}_{\{--+\},\{--+\}}^{(2,0),[\cdot]}$ and $\mathscr{M}_{\{--++\},\{--++\}}^{(3,0),[\cdot]}$, respectively,
 and once they are computed, they should be added to $\mathbf{M}^{(1,0)}$, $\mathbf{M}^{(2,0)}$ and $\mathbf{M}^{(3,0)}$  accordingly.
When the traversals of all the four trees are completed, the matrices $\mathbf{M}^{(k,0)}$, $k = 1,2,3$ are automatically built, 
 and we can then proceed to the computation of the reduced propagator $\mathbf{U}^{(k,0)}$ by \cref{Ur0,Ur0_truncated}.
Note that $\mathbf{M}^{(k+1,1)}$ for $k=1,2,3$ can also be constructed via depth-first tree traversal
because they also have similar recurrence relations as $\mathbf{M}^{(k,0)}$.

For general $\dk$,
 the implementation of t-SMatPI requires more memory than SMatPI.
In SMatPI, 
 we only need to store the values of kernel matrices $\mathbf{M}^{(k,0)}$
 for $k=1,\dots,\dk$ so that 
 the memory cost grows linearly 
 with respect to $\dk$.
The t-SMatPI algorithm requires to record the
 memory kernel matrices $\M{k}{0}$, which is of order $O(\dk)$.
In addition, it demands memory for all the information from the root node to a certain leaf node, which requires $O(\dk^2)$ memory
 because one node on the $k$th level contains $k-1$ complex values.
Therefore,
 the total memory cost for t-SMatPI is $O(\dk^2)$.
Nevertheless, this is a smaller magnitude compared with the memory requirement $O(4^{\dk})$ in the i-QuAPI
and is generally affordable.

The computational cost of the t-SMatPI is also low.
The algorithm requires visits to all the nodes in the forest of four trees.
Each of the $4^k$ nodes in the $k$th level includes $O(k)$ quantities because the node represents $\Mscr{k}{0}{n}$ for $n=0,\dots,k-1$.
The computation of each $\Mscr{k}{0}{n}$ requires $O(k)$ operations based on \cref{eq_Mscr_recurrence}.
If we use an addition accumulator and a multiply accumulator to represent the summation and product terms in \cref{eq_Mscr_recurrence},
 this computational cost becomes $O(1)$.
Specifically, we have
\begin{equation}
\label{eq_sum_mul_accumulator}
\begin{split}
    &\sum_{j=0}^{n-1} \left(
    \F{k}{j}-1
    \right)\Mscr{k}{0}{j}
    = \sum_{j=0}^{n-2} \left(
    \F{k}{j}-1
    \right)\Mscr{k}{0}{j}
    + \left(\F{k}{n-1}-1\right) \Mscr{k}{0}{n-1} \\
    &\prod_{l=n+1}^{k-1} \F{k+1}{l}
    = \frac{\prod_{l=n}^{k-1} \F{k+1}{l}}{\F{k+1}{n}}
\end{split}
\end{equation}

Therefore, the total computational cost
 for a complete tree traversal is
\begin{equation*}
    \sum_{k=0}^{\dk} k 4^k 
    = \frac{4}{9} \left( 1-4^{\dk} + 3\times \dk 4^{\dk} \right)
    = O(\dk 4^{\dk}).
\end{equation*}
The computational complexity of the whole SMatPI algorithm is of the same order as the cost for preparing the SMatPI matrices. 
Therefore, the total computational cost of t-SMatPI is $O(\dk 4^{\dk})$.
This complexity
will be verified by our numerical experiments
 in \cref{sec_numerical_results}.
As we can see, the computational cost of t-SMatPI is of the same order as SMatPI.
However, the t-SMatPI reveals more properties
 behind the kernel matrices $\mathbf{M}^{(k,0)}$ ($\mathbf{M}^{(k+1,1)}$) because it directly utilizes the definition of SMatPI matrices.

To end this section,
 we write down the pseudocode in \cref{algo_t_SMatPI}
 for the construction of matrices $\mathbf{M}^{(k,0)},\mathbf{M}^{(k+1,1)}$ for $k=1,\dots,\dk$
 for general choose of $\dk$.

\begin{algorithm}[ht]
  \caption{t-SMatPI algorithm}\label{algo_t_SMatPI}
  \begin{algorithmic}[1]
  \medskip
    \State Set $\mathbf{M}^{(k,0)} \gets \mathbf{O}$,
    $\mathbf{M}^{(k+1,1)} \gets \mathbf{O}$
    for $k=1,\dots,\dk$.
    \medskip
    \For{$j$ from 0 to 1}
        \medskip
        \For{$\mathcal{T}$ in $\{\mathcal{T}_{\{-1\},\{-1\}},\mathcal{T}_{\{-1\},\{+1\}},\mathcal{T}_{\{+1\},\{-1\}},\mathcal{T}_{\{+1\},\{+1\}}\}$} \Comment{$\mathcal{T}_{\sigma_j}$ denotes the tree with root $\sigma_j$}
            \medskip
            \State $\mathcal{N} \gets$ the root node of $\mathcal{T}$;
            \medskip
            \While {$\mathcal{N}$ is not the last node of $\mathcal{T}$ based on depth-first search}
                \medskip
                \State {$\mathcal{N} \gets$ the next node in $\mathcal{T}$;}
                \medskip
                \State $k \gets$ the depth of the node $\mathcal{N}$;
                \medskip
                \If {$k=1$}
                    \medskip
                    \State Compute $\Mcal{1+j}{j}$ by \cref{eq_initial_condition} or \cref{eq_initial_21};
                    \medskip
                    \State $\M{1+j}{j} \gets \M{1+j}{j} + \Mcal{1+j}{j}$;
                \Else
                \If {$k=2$}
                    \medskip
                    \State Compute $\Mscr{2+j}{j}{j}$ by \cref{eq_Mscr20_Mcal10} or \cref{eq_M21_M31};
                    \medskip
                \Else
                    \medskip
                    \State Compute the addition accumulator and multiply accumulator by \cref{eq_sum_mul_accumulator};
                    \State Compute $\Mscr{k+j}{j}{l}$ by \cref{eq_Mscr_recurrence}
                    or \cref{eq_Mscr_recurrence_1}
                    for $l=j,j+1,\dots,k-2+j$;
                    \medskip
                \EndIf
                \medskip
                \State $\M{k+j}{j} \gets \M{k+j}{j} + (\F{k+j}{l}-1) \Mscr{k+j}{j}{l}$ for $l=j,j+1,\dots,k-2+j$;
                \medskip
                \EndIf
                \medskip
            \EndWhile
            \medskip
        \EndFor
        \medskip
    \EndFor
    \medskip
    \State Compute propagators $\mathbf{U}^{(k,0)}$ for $k=1,\cdots,N$ by \cref{Ur0,Ur0_truncated};
    \end{algorithmic}
\end{algorithm}


\section{Numerical results}
\label{sec_numerical_results}
In our numerical tests, we choose the spectral density as the Ohmic spectral density \cite{makri1999linear} where
\begin{equation*}
    J(\omega) = \frac{\pi}{2} \sum_{j=1}^L \frac{c_j^2}{\omega_j} \delta(\omega-\omega_j)
\end{equation*}
with $\delta$ being the Dirac delta function and $\omega_j$ and $c_j$ are given by
\begin{equation*}
    \omega_j = -\omega_c \ln\left(1-\frac{j}{L} (1-\e^{-\omega_{\max}/\omega_c})\right), \quad
    c_j = \omega_j \sqrt{\frac{\xi\omega_c}{L}(1-\e^{-\omega_{\max}/\omega_c})},
    \quad j = 1,\dots,L
\end{equation*}
where $\xi$ represents the coupling intensity
and $L$ is the number of harmonic oscillators in the bath.

\begin{figure}
    \centering
    \begin{subfigure}{0.48\textwidth}
        \includegraphics[width = \textwidth]{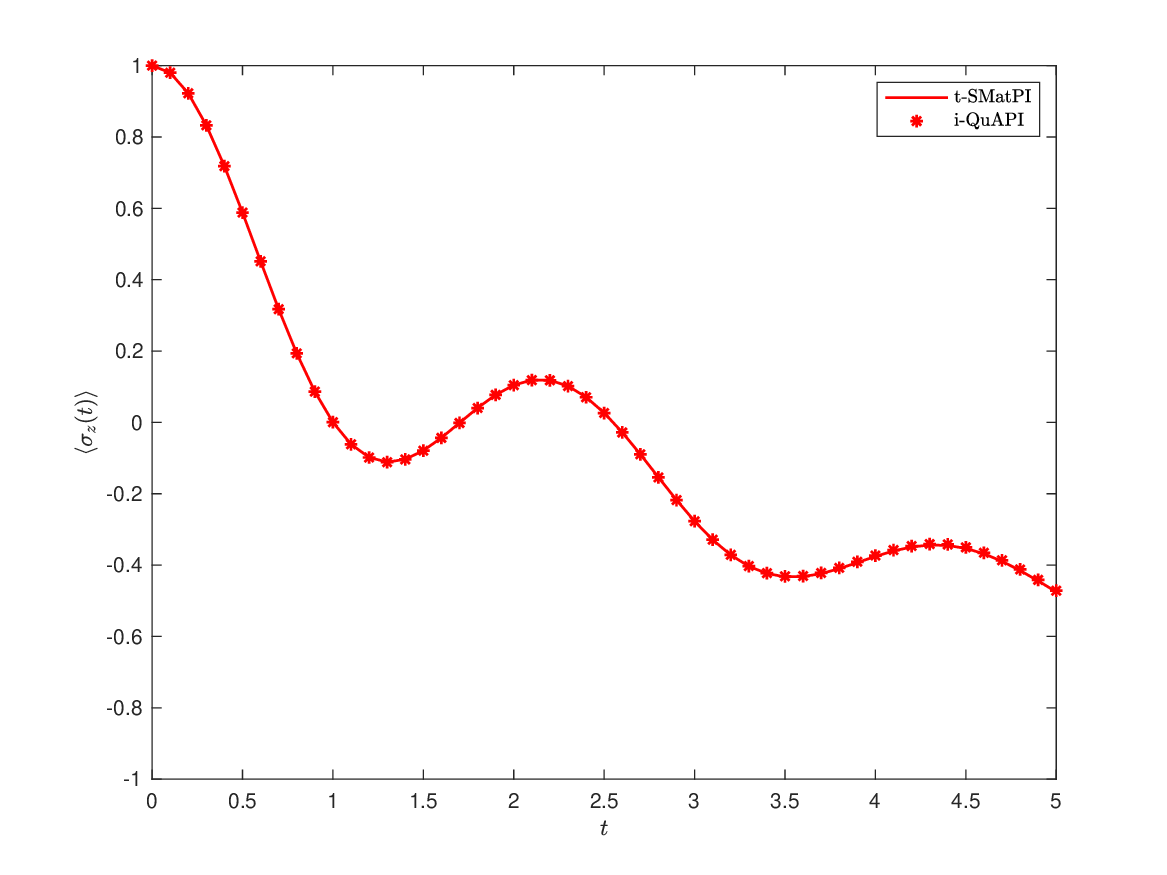}
        \caption{Numerical tests of t-SMatPI.}
        \label{fig_experiment_1}
    \end{subfigure}
    ~
    \begin{subfigure}{0.48\textwidth}
        \includegraphics[width = \textwidth]{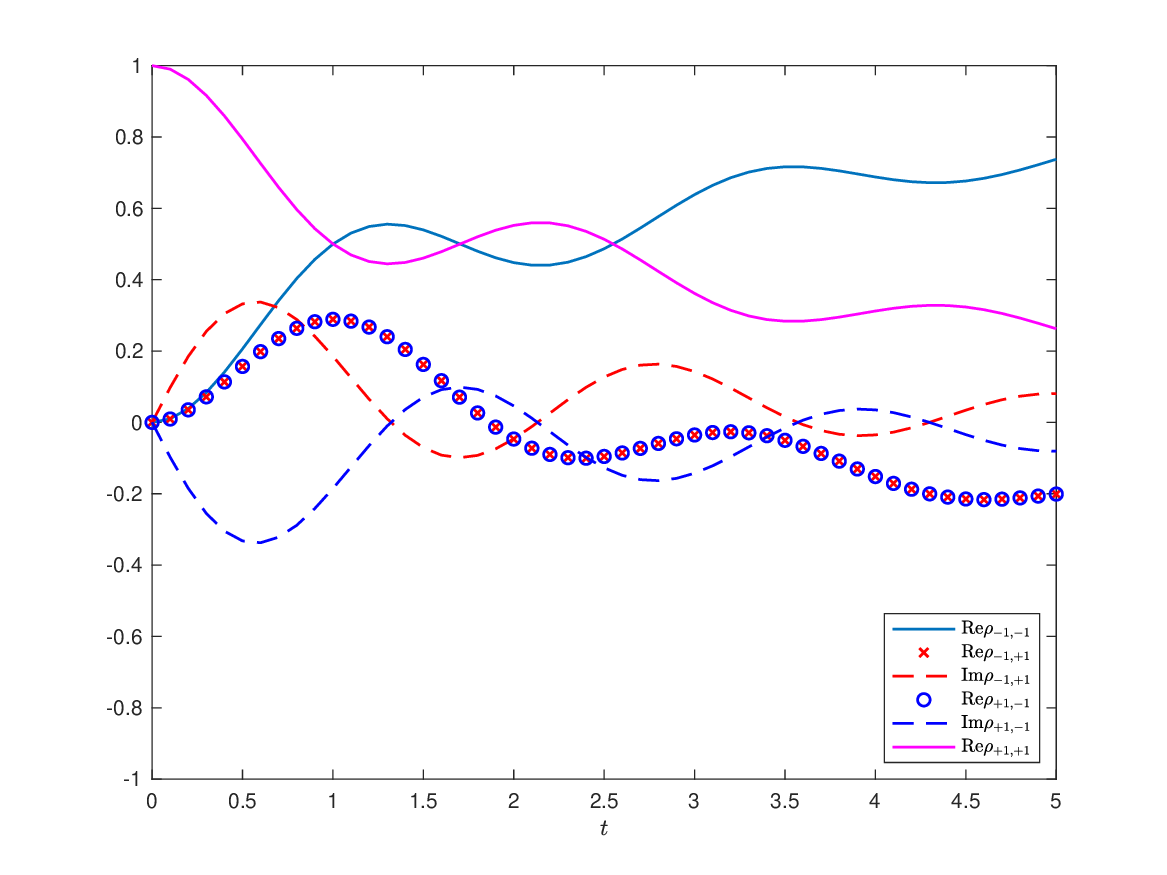}
        \caption{Evolution of the density matrix.}
        \label{fig_density}
    \end{subfigure}
    \caption{Numerical test result.}
\end{figure}

As our method computes exactly the same quantities as the SMatPI and the i-QuAPI,
 all the three algorithms should in principle yield the same results for the same parameter settings, up to the machine precision.
To show the validity of our algorithm,
 we use the following parameters
\begin{equation*}
    L = 400, \quad \Delta = 1, \quad \omega_c = 2.5, \quad \beta = 5, \quad \epsilon = 1, \quad \omega_{\max} = 10, \quad \xi = 0.2.
\end{equation*}
As for the numerical parameters,
 we choose $\dk = 10,$ and $\dt = 0.1$ in both i-QuAPI and our algorithm.
The results of the expectation of the observable $\expval{\sigma_z(t)}$ is given in
\cref{fig_experiment_1}.
From the result, it can be seen that our method computes exactly the same variables as i-QuAPI, which validates the t-SMatPI method and our implementation.


A few other properties can be observed from the evolution of the density matrix plotted in \cref{fig_density}.
From the figure, we notice that $\rho_{+1,-1}(t) = \overline{\rho_{-1,+1}(t)}$ 
where $\overline{\rho_{-1,+1}(t)}$ 
 denotes the complex conjugate of $\rho_{-1,+1}(t)$,
 since the real parts of $\rho_{+1,-1}(t)$ and $\rho_{-1,+1}(t)$
 are identical and their imaginary parts are opposite numbers for all $t$.
It can also be observed that the trace of the density matrix $\rho_{+1,+1}+\rho_{-1,-1}$ remains to be $1$ all the time,
 since the two lines are symmetric about the straight line $y = 0.5$.

\begin{figure}
    \centering
    \includegraphics[width = 0.5\textwidth]{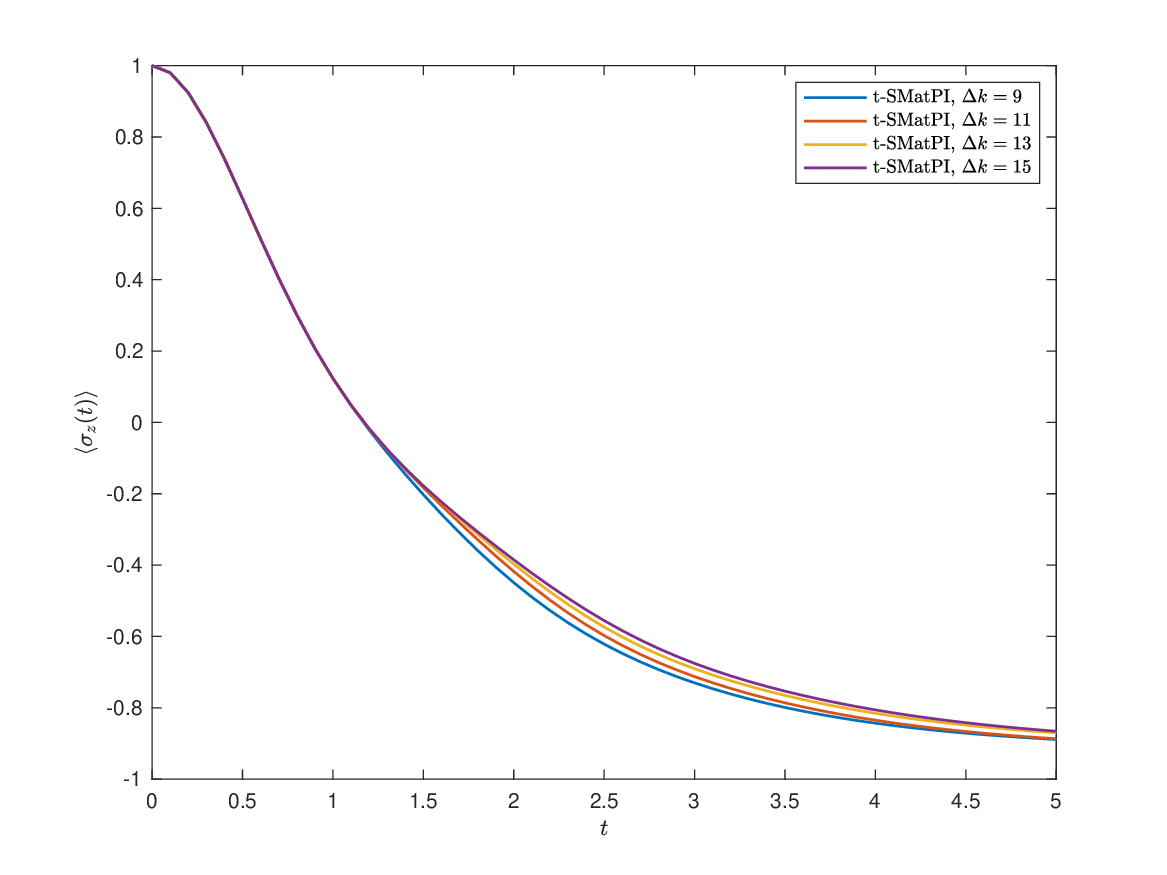}
    \caption{Numerical verification of the convergence of our method with respect to $\dk$.}
    \label{fig_experiment_2_dk}
\end{figure}

We then increase the coupling intensity $\xi$ to 1
 so that the system and the bath have a stronger coupling.
\Cref{fig_experiment_2_dk} shows the convergence of t-SMatPI
 with respect to $\dk$ for a fixed time step $\dt = 0.1$.
As the coupling between the system and the bath increases,
 the memory length should also increase to capture
 the influence functional accurately,
 and thus a larger $\dk$ is required 
 to ensure the accuracy.

\begin{figure}
    \centering
    \includegraphics[width = 0.5\textwidth]{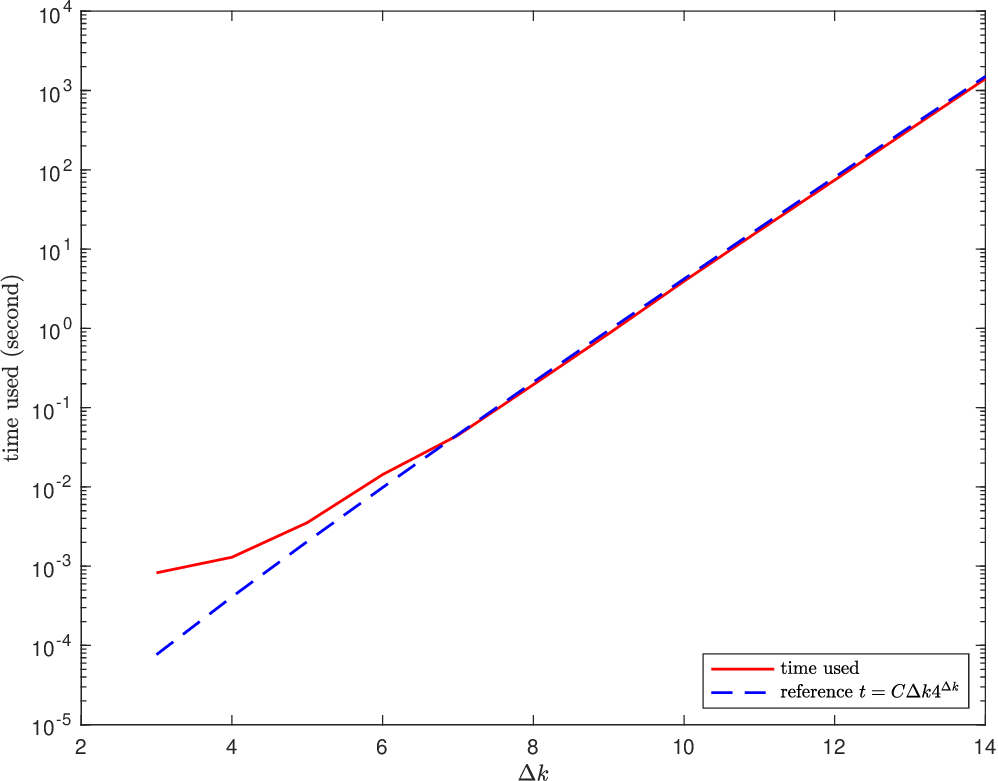}
    \caption{The growth of computational time with respect to $\dk$.}
    \label{fig_computational_time}
\end{figure}

The main improvement in our method is the reduction 
 of the computational cost.
We therefore choose different $\dk$ in our implementation and check the growth of computational time.
The following experiments 
 are run on the CPU model ``Intel\textsuperscript{\textregistered} Core\texttrademark \ i5-8257U'', 
 and no parallel computation is applied.
The results are shown in \cref{fig_computational_time}.
From the plot,
 it can be concluded that the computational complexity
 of our method is about $O(\dk 4^{\dk})$,
 consistent with our analysis in \cref{sec_implementation}.
Compared to the original SMatPI method, our algorithm allows us to choose a larger memory length $\dk$ when necessary.

\section{Conclusion and future work}
\label{sec_conclusion}
In this paper,
 we briefly introduce the SMatPI method
 in the simulation of system-bath dynamics
 and analyze its time complexity.
Utilizing the special structure of the diagrammatic representation,
 we derive a recurrence relation
 between memory kernels of adjacent time steps 
 for a fixed path.
A special tree traversal algorithm (t-SMatPI) is then developed, allowing us to 
maintain low computational cost by the recurrence relation.
Our study is currently restricted to the spin-boson model,
 but the t-SMatPI method can be generalized naturally to more complicated models such as the Feymann path amplitudes \cite{makri2021small1} and the spin chain models \cite{makri2021small4}.
More applications will studied in our future works.

At the end of this paper, we would like to discuss possible inspirations from the t-SMatPI method. 
Although our work generates exactly the same result as the original SMatPI method (regardless of round-off error), it provides a perspective to understand the memory kernel in the open quantum system.
In the SMatPI method, the computation of the matrices $\mathbf{M}^{(k+j,j)}$ is equivalent to the computation of the kernel function $\mathcal{K}(\cdot)$ in the following Nakajima-Zwanzig equation \cite{nakajima1958quantum,zwanzig1960ensemble}:
\begin{equation}
\label{eq_nakajima_zwanzig}
\frac{\mathrm{d}\rho_s}{\mathrm{d}t} = -\ii [H_s, \rho_s] + \int_0^t \mathcal{K}(\tau) \rho_s(t-\tau) \,\mathrm{d}\tau.
\end{equation}
The function $\mathcal{K}(\cdot)$ is usually expressed using projection operators \cite{nakajima1958quantum,zwanzig1960ensemble,smirne2010nakajima,xu2018convergence}.
But our study may help uncover an explicit and computable formula for the kernel function. This can be done by taking the limit $\Delta t \rightarrow 0$ and finding a continuous formulation of the matrices $\mathbf{M}^{(j+k,j)}$. A similar approach has been adopted in \cite{wang2022differential}. Here we will briefly demonstrate the general idea of this strategy. 

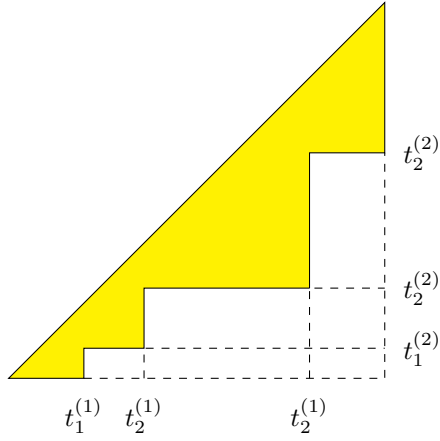
\begin{figure}
    \centering
    \begin{tikzpicture}
        \draw [-, fill=yellow] (4,1.2) -- (4,3) -- (5,3) -- (5,5) -- (0,0) -- (1,0) -- (1,0.4) -- (1.8,0.4) -- (1.8,1.2) -- (4,1.2);
        \node at (1,-0.5) {$t_1^{(1)}$};
        \node at (1.8,-0.5) {$t_2^{(1)}$};
        \node at (4,-0.5) {$t_{2}^{(1)}$};
        \node at (5.5,0.4) {$t_1^{(2)}$};
        \node at (5.5,1.2) {$t_2^{(2)}$};
        \node at (5.5,3) {$t_{2}^{(2)}$};
        \draw [dashed] (1.8,0) -- (1.8, 0.4) -- (5,0.4);
        \draw [dashed] (4,0) -- (4, 1.2) -- (5,1.2);
        \draw [dashed] (1,0) -- (5,0) -- (5,3);
    \end{tikzpicture}
    \caption{A staircase-like shape}
    \label{fig_staircase}
\end{figure}

Following \cite{wang2022differential}, we consider the continuous limit of the influence functional
\begin{equation*}
    \mathcal{F}(\sigma_1,\sigma_2,\tau)
    = -(\sigma_1^+ - \sigma_1^-)
    (\tilde{\eta}(\tau)\sigma_2^+ - {\overline{\tilde{\eta}(\tau)}} \sigma_2^-),
\end{equation*}
and it holds that
\begin{equation*}
    \mathcal{F}(\sigma_1,\sigma_2,\tau) 
     = \lim_{\dt\rightarrow 0}\frac{\F{j_1}{j_2} - 1}{\dt^2}.
\end{equation*}
if $\sigma_{j_1}^\pm = \sigma_1$, $\sigma_{j_2}^\pm = \sigma_2$ and $\tau=(j_1-j_2)\dt$.
In the discrete scheme, $\F{j_1}{j_2}$ are multiplied on a staircase-like shape
and the quantities $\F{j_1}{j_2}-1$ is multiplied for each step of a staircase.
Similarly, for any fixed path $g$ (refer to \cite{wang2022differential} for more details), an integral is required in the continuous limit
\begin{equation}
    \mathcal{B}(g,\mathcal{S})=
    \exp\left(
    \int_{\mathcal{S}}
    \mathcal{F}\left(g(x_1),g(x_2),x_1-x_2)\right)
    \dd x_1 \dd x_2
    \right)
    \left(
    \prod_{k=1}^{d+1}
    \mathcal{F}\left(g(t_k^{(1)}),g(t_{k-1}^{(2)}),t_k^{(1)}-t_{k-1}^{(2)} \right)
    \right)
\end{equation}
where $\mathcal{S}$ represents a staircase-like shape like \cref{fig_staircase}.
The contribution from the system part is:
\begin{equation}
\begin{split}
    \mathcal{Y}(g)
    &= \e^{-\ii\left(\mel{r_0^+}{H_0}{r_0^+} - \mel{r_0^-}{H_0}{r_0^-}\right)\tau_1} \\
    &\times
    \prod_{j=1}^{D} \left[
    \e^{-\ii\left(\mel{r_j^+}{H_0}{r_j^+} - \mel{r_j^-}{H_0}{r_j^-}\right)\tau_{j+1}}
    \left(
    -\ii \mel{r_j^+}{H_0}{r_{j-1}^+} \delta_{\sgn_j}^+
    + \ii \mel{r_{j-1}^-}{H_0}{r_j^-} \delta_{\sgn_j}^-
    \right)
    \right].
\end{split}
\end{equation}
The SMatPI matrices in continuous form is then given by
\begin{equation}
\label{eq:M(t)}
    M(t) = 
    \int_{g\in\mathscr{P}}
    \int_{\mathcal{S}\in\mathscr{S}}
    \mathcal{B}(g,\mathcal{S}) \mathcal{Y}(g)
\end{equation}
where $\mathscr{S}$ is the set of all possible staircase-like shapes and $\mathscr{P}$ is the set of all possible paths. 
This function $M(t)$ is the continuous formulation of $\mathrm{M}$ in the discrete form.
Based on \cref{Ur0},
we have
\begin{equation*}
    U(t) = \int_{0}^t M(t-\tau) U(\tau) \dd \tau
\end{equation*}
and
\begin{equation}
    \dv{U(t)}{t} = M(0) U(t) 
    + \int_0^t \dv{M(t-\tau)}{t} U(\tau) \dd \tau
\end{equation}
By \cref{eq_density}, we see that $M'(t)$ plays the same role as $\mathcal{K}(t)$ in the Nakajima-Zwanzig equation \eqref{eq_nakajima_zwanzig}.

The computation of the kernel in the t-SMatPI method can be considered as a discretization of \eqref{eq:M(t)},
 and the accuracy is expected to be second order in $\Delta t$.
In our future work, we will further explore the implication of the recurrence relations \eqref{eq_initial_21}--\eqref{eq_split_M1_Mscr} in the continuous sense. It is likely that their continuous limit will be integro-differential equations, which may reveal some new properties of the memory kernel.
Once the equations are formulated, classical numerical techniques might be applied to obtain higher-order numerical schemes to compute the memory kernel, resulting in more efficient or more accurate methods to study the system-bath dynamics.


\section*{Appendix. Catalan's triangle and its properties}
\label{sec_catlan_triangle}
In combinatorics, the sequence of Catalan numbers is closely related to
 the Catalan's triangle in \cref{fig_catalan_triangle} \cite{shapiro1976catalan}.
The $k$th number in the $n$th row 
 ($0 \leqslant k\leqslant n$, $n\geqslant 0$) 
 in the Catalan's triangle is given by
\begin{equation*}
    T(n,k) = \binom{n+k}{k} - \binom{n+k}{k-1}
    = \frac{n-k+1}{n+1}\binom{n+k}{k}.
\end{equation*}
\begin{figure}
\begin{center}
\begin{tabular}{c|c c c c c c}
    & $k=0$ & $k=1$ & $k=2$ & $k=3$ & $k=4$ & $\dots$ \\ \hline
    $n=0$ & 1 &  \\
    $n=1$ & 1 & 1  \\
    $n=2$ & 1 & 2 & 2 \\
    $n=3$ & 1 & 3 & 5 & 5 \\
    $n=4$ & 1 & 4 & 9 & 14 & 14 \\
    $\vdots$ & $\vdots$ & $\vdots$ & $\vdots$ & $\vdots$ & $\vdots$ & $\ddots$
\end{tabular}
\end{center}
    \caption{First five rows of Catalan's Triangle $T(n,k)$.}
    \label{fig_catalan_triangle}
\end{figure}
In this paper,
 we utilize the following properties of Catalan's triangle:
\begin{itemize}
    \item[(\romannumeral 1)] the last number in each row of Catalan's triangle
    is the Catalan number.
    \begin{equation}
        \label{eq_CnTnn}
        C_n = T(n,n).
    \end{equation}
    \item[(\romannumeral 2)] the sum of the first $k$ numbers in the $n$ row
    gives us $T(n+1,k)$.
    \begin{equation}
        \label{eq_recurrence}
        T(n+1,k) = \sum_{j=1}^{k} T(n,j)
    \end{equation}
    In particular, the sum of each row in the Catalan's triangle
     is the sequence of Catalan numbers.
    \begin{equation}
        \label{eq_sumTnj}
        C_{n+1} = \sum_{j=1}^n T(n,j).
    \end{equation}
    \item[(\romannumeral 3)] $T(n,k)$ represents the number of Dyck paths from $(0,0)$ to $(n+1,n+1)$ that arriving $x=n+1$ at the point $(n+1,k)$.
\end{itemize}

\bibliographystyle{abbrv}
\bibliography{myBib.bib}

\begin{thebibliography}{10}

\bibitem{boag2018inclusion}
A.~Boag, E.~Gull, and G.~Cohen.
\newblock Inclusion-exclusion principle for many-body diagrammatics.
\newblock {\em Phys. Rev. B}, 98(11):115152, 2018.

\bibitem{buser2017initial}
M.~Buser, J.~Cerrillo, G.~Schaller, and J.~Cao.
\newblock Initial system-environment correlations via the transfer-tensor
  method.
\newblock {\em Phys. Rev. A}, 96(6):062122, 2017.

\bibitem{cai2020inchworm}
Z.~Cai, J.~Lu, and S.~Yang.
\newblock Inchworm {M}onte {C}arlo method for open quantum systems.
\newblock {\em Commun. Pure Appl. Math.}, 73(11):2430--2472, 2020.

\bibitem{cai2022fast}
Z.~Cai, J.~Lu, and S.~Yang.
\newblock Fast algorithms of bath calculations in simulations of quantum
  system-bath dynamics.
\newblock {\em Comput. Phys. Commun.}, page 108417, 2022.

\bibitem{cai2022numerical}
Z.~Cai, J.~Lu, and S.~Yang.
\newblock Numerical analysis for inchworm {M}onte {C}arlo method: sign problem
  and error growth.
\newblock {\em Math. Comput.}, 2022.

\bibitem{cai2023bold}
Z.~Cai, G.~Wang, and S.~Yang.
\newblock The bold-thin-bold diagrammatic {M}onte {C}arlo method for open
  quantum systems.
\newblock {\em SIAM J. Sci. Comput.}, 45(4):A1812--A1843, 2023.

\bibitem{cerrillo2014nonmarkovian}
J.~Cerrillo and J.~Cao.
\newblock Non-{M}arkovian dynamical maps: numerical processing of open quantum
  trajectories.
\newblock {\em Phys. Rev. Lett.}, 112(11):110401, 2014.

\bibitem{chen2017inchworm1}
H.-T. Chen, G.~Cohen, and D.~R. Reichman.
\newblock Inchworm {M}onte {C}arlo for exact non-adiabatic dynamics. i. theory
  and algorithms.
\newblock {\em J. Chem. Phys.}, 146(5):054105, 2017.

\bibitem{chen2020non}
Y.-Q. Chen, K.-L. Ma, Y.-C. Zheng, J.~Allcock, S.~Zhang, and C.-Y. Hsieh.
\newblock Non-{M}arkovian noise characterization with the transfer tensor
  method.
\newblock {\em Phys. Rev. Appl.}, 13(3):034045, 2020.

\bibitem{deutsch1999dyck}
E.~Deutsch.
\newblock {D}yck path enumeration.
\newblock {\em Discrete Math.}, 204(1-3):167--202, 1999.

\bibitem{feynman1963theory}
R.~P. Feynman and F.~Vernon.
\newblock The theory of a general quantum system interacting with a linear
  dissipative system.
\newblock {\em Ann. Phys.}, 24:118--173, 1963.

\bibitem{harris2008combinatorics}
J.~Harris, J.~Hirst, and M.~Mossinghoff.
\newblock {\em Combinatorics and Graph Theory}.
\newblock Undergraduate Texts in Mathematics. Springer New York, 2008.

\bibitem{loh1990sign}
L.~E. Jr., J.~Gubernatis, R.~Scalettar, S.~White, D.~Scalapino, and R.~Sugar.
\newblock Sign problem in the numerical simulation of many-electron systems.
\newblock {\em Phys. Rev. B}, 41(13):9301, 1990.

\bibitem{kundu2023pathsum}
S.~Kundu and N.~Makri.
\newblock Pathsum: A {C}++ and fortran suite of fully quantum mechanical
  real-time path integral methods for (multi-) system+ bath dynamics.
\newblock {\em J. Chem. Phys.}, 158(22), 2023.

\bibitem{makri1992improved}
N.~Makri.
\newblock Improved {F}eynman propagators on a grid and non-adiabatic
  corrections within the path integral framework.
\newblock {\em Chem. Phys. Lett.}, 193(5):435--445, 1992.

\bibitem{makri1995numerical}
N.~Makri.
\newblock Numerical path integral techniques for long time dynamics of quantum
  dissipative systems.
\newblock {\em J. Math. Phys.}, 36(5):2430--2457, 1995.

\bibitem{makri1998quantum}
N.~Makri.
\newblock Quantum dissipative dynamics: A numerically exact methodology.
\newblock {\em J. Phys. Chem. A}, 102(24):4414--4427, 1998.

\bibitem{makri1999linear}
N.~Makri.
\newblock The linear response approximation and its lowest order corrections:
  An influence functional approach.
\newblock {\em J. Phys. Chem. B}, 103(15):2823--2829, 1999.

\bibitem{makri2014blip}
N.~Makri.
\newblock Blip decomposition of the path integral: Exponential acceleration of
  real-time calculations on quantum dissipative systems.
\newblock {\em J. Chem. Phys.}, 141(13):134117, 2014.

\bibitem{makri2016blip}
N.~Makri.
\newblock Blip-summed quantum--classical path integral with cumulative quantum
  memory.
\newblock {\em Faraday Discuss.}, 195:81--92, 2016.

\bibitem{makri2020small2}
N.~Makri.
\newblock Small matrix disentanglement of the path integral: overcoming the
  exponential tensor scaling with memory length.
\newblock {\em J. Chem. Phys.}, 152(4), 2020.

\bibitem{makri2020small1}
N.~Makri.
\newblock Small matrix path integral for system-bath dynamics.
\newblock {\em J. Chem. Theory Comput.}, 16(7):4038--4049, 2020.

\bibitem{makri2021small1}
N.~Makri.
\newblock Small matrix decomposition of {F}eynman path amplitudes.
\newblock {\em J. Chem. Theory Comput.}, 17(7):3825--3829, 2021.

\bibitem{makri2021small4}
N.~Makri.
\newblock Small matrix modular path integral: iterative quantum dynamics in
  space and time.
\newblock {\em Phys. Chem. Chem. Phys.}, 23(22):12537--12540, 2021.

\bibitem{makri2021small2}
N.~Makri.
\newblock Small matrix path integral for driven dissipative dynamics.
\newblock {\em J. Phys. Chem. A}, 125(48):10500--10506, 2021.

\bibitem{makri2021small3}
N.~Makri.
\newblock Small matrix path integral with extended memory.
\newblock {\em J. Chem. Theory Comput.}, 17(1):1--6, 2021.

\bibitem{makri1995tensor1}
N.~Makri and D.~E. Makarov.
\newblock Tensor propagator for iterative quantum time evolution of reduced
  density matrices. i. theory.
\newblock {\em J. Chem. Phys.}, 102(11):4600--4610, 1995.

\bibitem{makri1995tensor2}
N.~Makri and D.~E. Makarov.
\newblock Tensor propagator for iterative quantum time evolution of reduced
  density matrices. ii. numerical methodology.
\newblock {\em J. Chem. Phys.}, 102(11):4611--4618, 1995.

\bibitem{nakajima1958quantum}
S.~Nakajima.
\newblock On quantum theory of transport phenomena: Steady diffusion.
\newblock {\em Prog. Theor. Phys.}, 20(6):948--959, 1958.

\bibitem{prokof1998polaron}
N.~V. Prokof'ev and B.~V. Svistunov.
\newblock Polaron problem by diagrammatic quantum {M}onte {C}arlo.
\newblock {\em Phys. Rev. Lett.}, 81(12):2514, 1998.

\bibitem{prokof2007bold}
N.~Prokof’ev and B.~Svistunov.
\newblock Bold diagrammatic {M}onte {C}arlo technique: When the sign problem is
  welcome.
\newblock {\em Phys. Rev. Lett.}, 99(25):250201, 2007.

\bibitem{shapiro1976catalan}
L.~W. Shapiro.
\newblock A {C}atalan triangle.
\newblock {\em Discrete Math.}, 14(1):83--90, 1976.

\bibitem{shi2003new}
Q.~Shi and E.~Geva.
\newblock A new approach to calculating the memory kernel of the generalized
  quantum master equation for an arbitrary system--bath coupling.
\newblock {\em J. Chem. Phys.}, 119(23):12063--12076, 2003.

\bibitem{smirne2010nakajima}
A.~Smirne and B.~Vacchini.
\newblock {N}akajima-{Z}wanzig versus time-convolutionless master equation for
  the non-{M}arkovian dynamics of a two-level system.
\newblock {\em Phys. Rev. A}, 82(2):022110, 2010.

\bibitem{wang2022differential}
G.~Wang and Z.~Cai.
\newblock Differential equation based path integral for open quantum systems.
\newblock {\em SIAM J. Sci. Comput.}, 44(3):B771--B804, 2022.

\bibitem{wang2023real}
G.~Wang and Z.~Cai.
\newblock Real-time simulation of open quantum spin chains with the inchworm
  method.
\newblock {\em J. Chem. Theory Comput.}, 2023.

\bibitem{werner2009diagrammatic}
P.~Werner, T.~Oka, and A.~J. Millis.
\newblock Diagrammatic {M}onte {C}arlo simulation of nonequilibrium systems.
\newblock {\em Phys. Rev. B}, 79(3):035320, 2009.

\bibitem{xu2018convergence}
M.~Xu, Y.~Yan, Y.~Liu, and Q.~Shi.
\newblock Convergence of high order memory kernels in the nakajima-zwanzig
  generalized master equation and rate constants: Case study of the spin-boson
  model.
\newblock {\em J. Chem. Phys.}, 148(16), 2018.

\bibitem{yang2021inclusion}
S.~Yang, Z.~Cai, and J.~Lu.
\newblock Inclusion--exclusion principle for open quantum systems with bosonic
  bath.
\newblock {\em New J. Phys.}, 23(6):063049, 2021.

\bibitem{zhang2006nonequilibrium}
M.-L. Zhang, B.~J. Ka, and E.~Geva.
\newblock Nonequilibrium quantum dynamics in the condensed phase via the
  generalized quantum master equation.
\newblock {\em J. Chem. Phys.}, 125(4):044106, 2006.

\bibitem{zwanzig1960ensemble}
R.~Zwanzig.
\newblock Ensemble method in the theory of irreversibility.
\newblock {\em J. Chem. Phys.}, 33(5):1338--1341, 1960.

\end{thebibliography}
\end{document}